\documentclass[
reprint,
superscriptaddress,
preprintnumbers,
 amsmath,
 amssymb,
 prc,
]{revtex4-1}

\usepackage{graphicx}
\usepackage{dcolumn}
\usepackage{bm}
\usepackage[colorlinks,
linkcolor=blue,
anchorcolor=blue,
citecolor=blue,
urlcolor=blue]{hyperref}
\usepackage{xcolor}
\usepackage{soul}
\usepackage{ulem}

\newcommand{\MSbar}{\overline{\mathrm{MS}}}
\newcommand{\Lmu}{\ln\!\frac{Q^2}{\mu^2}}
\newcommand{\as}{\alpha_s(Q^2)}
\begin{document}


\title{Parton Fragmentation Functions Extracted with a Physics-Informed Neural Network}

\author{Si-Wei Dai}
\email[]{swdai@mails.ccnu.edu.cn}
\affiliation{Key Laboratory of Quark and Lepton Physics (MOE) \& Institute of Particle Physics, Central China Normal University, Wuhan 430079, China}
\affiliation{Artificial Intelligence and Computational Physics Research Center, Central China Normal University, Wuhan 430079, China}
\author{Fu-Peng Li}
\email[]{fpli@fudan.edu.cn}
\affiliation{Key Laboratory of Nuclear Physics and Ion-beam Application (MOE) \& Institute of Modern Physics, Fudan University, Shanghai 200433, China}
\affiliation{Shanghai Research Center for Theoretical Nuclear Physics,
NSFC and Fudan University, Shanghai 200438, China}
\affiliation{Key Laboratory of Quark and Lepton Physics (MOE) \& Institute of Particle Physics, Central China Normal University, Wuhan 430079, China}
\affiliation{Artificial Intelligence and Computational Physics Research Center, Central China Normal University, Wuhan 430079, China}

\author{Long-Gang Pang}
\email[]{lgpang@ccnu.edu.cn}
\affiliation{Key Laboratory of Quark and Lepton Physics (MOE) \& Institute of Particle Physics, Central China Normal University, Wuhan 430079, China}
\affiliation{Artificial Intelligence and Computational Physics Research Center, Central China Normal University, Wuhan 430079, China}

\author{Xin-Nian Wang}
\email[]{xnwang@lbl.gov}
\affiliation{Key Laboratory of Quark and Lepton Physics (MOE) \& Institute of Particle Physics, Central China Normal University, Wuhan 430079, China}
\affiliation{Artificial Intelligence and Computational Physics Research Center, Central China Normal University, Wuhan 430079, China}

\author{Ben-Wei Zhang}
\email[]{bwzhang@ccnu.edu.cn}

\author{Han-Zhong Zhang}
\email[]{zhanghz@ccnu.edu.cn}
\affiliation{Key Laboratory of Quark and Lepton Physics (MOE) \& Institute of Particle Physics, Central China Normal University, Wuhan 430079, China}
\affiliation{Artificial Intelligence and Computational Physics Research Center, Central China Normal University, Wuhan 430079, China}

\date{\today}
\begin{abstract}
Reliable predictions of many high-energy strong interaction processes rely heavily on the non-perturbative parton fragmentation functions (FFs) extracted from existing experimental data. Conventional methods often require parameterized forms of FFs and additional scale evolution according to the Dokshitzer-Gribov-Lipatov-Altarelli-Parisi (DGLAP) evolution equations.
We introduce a novel approach to determining parton FFs using a Physics-Informed Neural Network (PINN). Unlike traditional methods, our approach does not require prior parameterized forms and directly integrates the DGLAP evolution equations into the neural network architecture, allowing the FFs to automatically satisfy these equations. We present new sets of parton FFs extracted from hadron spectra in electron-positron annihilation processes at next-to-leading order (NLO) in pQCD using this new technique. To validate our approach, we calculate charged hadron spectra in proton-(anti)proton collisions using the extracted FFs and demonstrate that the results align well with experimental data across a large range of colliding energies ($\sqrt{s}$ = 130, 200, 500, 630, 900, 1800, 2760, 5020, 5440, 7000 GeV). Our findings indicate that the PINN method not only simplifies the extraction process but also enhances the universal applicability of FFs across different energy scales. By eliminating the need for parameterized forms and additional DGLAP evolution, our approach represents a significant step forward toward fast and accurate extractions of non-perturbative quantities such as parton fragmentations functions and parton distribution functions.
\end{abstract}


\maketitle


\section{INTRODUCTION}

Hard processes in high-energy collisions are characterized by large momentum transfers or short distance in space and time. While the asymptotic freedom of Quantum Chromodynamics (QCD) allows the perturbative calculation of partonic cross sections, the confinement of quarks and gluons necessitates a framework that connects these short-distance interactions to experimentally observed hadrons. This connection is provided by the factorization theorem~\cite{Collins:1989gx}, which separates the calculable hard scattering from universal non-perturbative quantities: Parton Distribution Functions (PDFs) for the initial state~\cite{Hou:2019qau,NNPDF:2021njg} and Fragmentation Functions (FFs) for the final state~\cite{Sato:2019yez,Bertone:2018ecm}. Within the factorization framework, PDFs and FFs are process-independent (universal): once determined in one process, e.g., deeply inelastic scattering (DIS) and single-inclusive electron-positron annihilation, the same PDFs and FFs, defined in the same factorization scheme and at the same scale, can be applied to other processes such as the Drell-Yan processes, the semi-inclusive deep-inelastic scattering and single-inclusive hadron production in $pp$ collisions. Consequently, precise knowledge of PDFs and FFs is indispensable for theoretical predictions for strong interaction processes in high-energy collisions.

Partons produced in hard interactions are initially off-shell, with virtualities of the order of the hard scale $Q$. These hard partons go through final-state radiation and reduce their virtuality before hadronization. This radiation-induced parton branching is the physical origin of the scale dependence of the fragmentation function $D_i^h(z,\mu)$ which encodes the number density of a hadron $h$ carrying a momentum fraction $z$ from the hadronization of a parton $i$ at factorization scale $\mu$. Although FFs are non-perturbative in origin, their scale dependence is governed by perturbative QCD and is described by the Dokshitzer–Gribov–Lipatov–Altarelli–Parisi (DGLAP) equations~\cite{Gribov:1972ri,Gribov:1972rt,Dokshitzer:1977sg,Altarelli:1977zs}. Crucially, these evolution equations are process-independent, so that FFs extracted at $\mu_0$ in one process can be consistently evolved—using the same splitting kernels—to the scales relevant for other processes.

As non-perturbative quantities defined on the light cone, FFs present a challenge for first-principles computations. While recent developments in Lattice QCD based on Large Momentum Effective Theory (LaMET)~\cite{Ji:2013dva,Ji:2014gla,Ji:2020ect} have successfully accessed light-cone observables (e.g., generalized parton distributions~\cite{Alexandrou:2020zbe,Bhattacharya:2022aob,Ding:2024saz}), the precise determination of FFs for phenomenological applications currently relies primarily on global QCD analyses of experimental data. The core experimental constraints are provided by data of single-inclusive electron-positron annihilation (SIA) from the ALEPH~\cite{ALEPH:1995njx,ALEPH:2003obs}, TASSO~\cite{TASSO:1990cdg}, TPC~\cite{TPCTwoGamma:1988yjh}, OPAL~\cite{OPAL:1998arz}, DELPHI~\cite{DELPHI:1998cgx}, and SLD~\cite{SLD:2003ogn} Collaborations. SIA provides the cleanest and most relevant data sets to access the FFs since its theoretical calculations do not require simultaneous knowledge of PDFs. However, it does not allow a complete flavor decomposition of quark and antiquark FFs along with a direct determination of the gluon FF. Therefore, it is often necessary to consider semi-inclusive deep-inelastic scattering (SIDIS)~\cite{COMPASS:2018lzp,COMPASS:2020oyf,HERMES:2012uyd,COMPASS:2016xvm} and the single-inclusive hadrons production in proton-(anti)proton collisions~\cite{ALICE:2013txf,ALICE:2014nqx,ALICE:2018hza,STAR:2003fka,PHENIX:2002diz,CMS:2011mry,CMS:2012aa,CMS:2016xef,CMS:2018yyx,ALICE:2020jsh,STAR:2013zyt,PHENIX:2015fxo}. In recent studies, nevertheless, only SIA data were included in the HKKS16~\cite{Hirai:2016loo}, JAM16~\cite{Sato:2016wqj}, and NNFF1.0~\cite{Bertone:2017tyb} analyses to extract the FFs for the light charged hadrons (i.e., $\pi^{\pm}, K^{\pm}, p/\bar{p}$). The analysis in Refs.~\cite{deFlorian:2014xna,deFlorian:2017lwf}, however, included both SIA and $pp$ data to extract the FFs for the light charged hadrons. The FFs of heavier hadrons were also studied in the analyses in Refs.~\cite{Salajegheh:2019ach,Benzke:2019usl,Delpasand:2020vlb} with  experimental data also mainly from SIA. 

A central challenge in the extraction of FFs is to balance parametric flexibility that is needed to capture non-perturbative dynamics with theoretical consistency in the scale dependence dictated by QCD evolution. Traditional fits often employ relatively rigid functional forms at an input scale $\mu_0$, which might induce parametrization bias and complicate optimization in high-dimensional settings. Neural-network parameterizations offer a flexible, non-parametric representation and have become widely used in high-energy physics~\cite{Pang:2016vdc,Boehnlein:2021eym,He:2023zin,Ma:2023zfj,Zhou:2023pti,Fernando:2025xzv}. In conventional neural network implementations, the optimization repeatedly calls numerical DGLAP evolution and the associated convolution in the cross section for hadron production, which can be computationally demanding when scanning large parameter spaces and kinematic regions.

To address these issues in this study, we implement a Physics-Informed Neural Network (PINN)~\cite{RAISSI2019686,9429985,DBLP:journals/corr/abs-2201-05624} that incorporates the DGLAP equations directly into the training processes. The network represents the multi-dimensional FFs $D_i^{h^{\pm}}(z,\mu)$ and is trained with a composite loss function comprising (1) a data term enforcing agreement with experimental data and (2) an evolution-residual term penalizing deviations from DGLAP evolution. We implement the evolution constraint in the Mellin space, where DGLAP convolutions reduce to algebraic products, improving numerical efficiency. Automatic differentiation~\cite{JMLR:v18:17-468} is used to compute the scale derivatives required by the evolution residual within the optimization loop. By enforcing the DGLAP evolution  during training, the extracted FFs will be globally consistent with QCD evolution across the $(z,\mu)$ domain within a controlled numerical tolerance while retaining the flexibility guaranteed by the universal approximation property of neural networks~\cite{HORNIK1989359}.

The remainder of this paper is organized as follows. In Sec.~\ref{TWO}, we review the theoretical framework for single-inclusive hadron production in electron-positron annihilation and the DGLAP evolution equations. Section~\ref{THREE} details the PINN methodology, the experimental data sets, and the error propagation strategy, followed by a presentation of the network's training performance. The main physical results of the extracted FFs are discussed in Sec.~\ref{FOUR}, starting with the verification of DGLAP constraints and comparisons with KRE and AKK08 results. We further demonstrate the robustness of our method via closure tests and illustrate its phenomenological utility by calculating single-inclusive hadronic spectra for $pp(\bar{p})$ collisions at RHIC and LHC. Finally, conclusions are drawn in Sec.~\ref{FIVE}.

\section{Fragmentation Functions in single-inclusive electron-positron annihilation}
\label{TWO}

In this work, we focus on FFs for unidentified charged hadrons ($h^{\pm}$)~\cite{Moffat:2021dji,Soleymaninia:2018uiv,Bertone:2018ecm,Kretzer:2000yf,Albino:2008fy,deFlorian:2007ekg}. Existing determinations typically follow one of two strategies. A common approach reconstructs $D_i^{h^{\pm}}$ by summing the FFs of identified hadrons (e.g., $\pi^{\pm}$, $K^{\pm}$, $p/\bar{p}$) and adding a residual component to account for heavier or rarer hadrons~\cite{Moffat:2021dji,Kretzer:2000yf,Albino:2008fy}. This procedure can propagate and accumulate uncertainties from each identified channel and introduces model dependence through the residual term. The alternative is to extract $h^\pm$ FFs by fitting directly the corresponding experimental data~\cite{Bertone:2018ecm}. We will adopt this direct-extraction approach in this study and restrict ourselves to SIA measurements, taking advantage of the theoretically simpler $e^+e^-$ initial state and avoiding the uncertainty propagation and model dependence of the residual inherent to the summation method.


To improve the constraining power within an SIA-only framework, we exploit additional SIA observables beyond inclusive spectra. Flavor-tagged measurements ($uds$-tag, $c$-tag, and $b$-tag) provide sensitivity to the decomposition into light-, charm-, and bottom-quark fragmentation contributions. Moreover, we include measurements of the inclusive longitudinal cross section $d\sigma_L/dz$ (related to the longitudinal structure function $F_L$). This observable isolates the longitudinal component from the total cross section. Crucially, unlike the single-inclusive total cross section which is non-zero at leading order (LO), $\sigma_L$ vanishes at LO and arises only at next-to-leading order (NLO) via hard gluon radiation. This perturbative behavior provides a direct and enhanced sensitivity to the gluon channel. The combination of inclusive, flavor-tagged, and longitudinal SIA data thus provides complementary constraints that mitigate the traditional limitations of SIA-only extractions, enabling a precise determination of $h^{\pm}$ FFs in a controlled theoretical framework.

In the framework of collinear factorization, we can separate the QCD cross section into  a convolution of a perturbative hard part and the non-perturbative parton distribution functions (PDFs) or parton fragmentation functions (FFs). In this work, we consider the single-inclusive electron-positron annihilation process,
\begin{equation} \label{SIA process}
    e^+ + e^- \to \gamma^*/Z^0 \to h^{\pm} + X\ .
\end{equation}
According to the collinear factorization theorem, the cross section of the above process can be written as,
\begin{equation} \label{SIA cross section}
    \sigma = \hat{\sigma} \otimes FFs\ ,
\end{equation}
where $\hat{\sigma}$ is the hard partonic cross section. The details of the computation of the SIA cross sections are provided in some studies available in the literature, and we refer the reader to Refs.~\cite{Binnewies:1995pt,Nason:1993xx,Binnewies:1994ju} for a review. 

Following the notations in Ref.~\cite{Nason:1993xx}, we rewrite Eq.~\eqref{SIA cross section} in terms of the structure functions of the $e^+e^-$ annihilation,
\begin{eqnarray}
    F^{h^{\pm}}(z,Q) &=& \frac{1}{\sigma_{tot}}\frac{d\sigma^{h^{\pm}}}{dz} \nonumber \\
    & = & \frac{1}{\sigma_{tot}} \sum_i  C_i(z,Q) \otimes D_i^{h^{\pm}}(z,Q)\ ,
    \label{dx cs}
\end{eqnarray}
where $z$ is the fractional momentum defined as $z = {2p_T^{h^{\pm}}}/{\sqrt{s}}$, $p_T^{h^{\pm}}$ is the transverse momentum of the final-state hadron, and $C_i(z,Q)$ is the coefficient function, with explicit expressions given in Appendix~\ref{app:coeff_functions}; $D_i^{h^{\pm}}(z,Q)$ is the fragmentation function of the hadron $h^{\pm}$ from a parton $i (u,d,s,\cdots,g)$, and $Q^2$ is the squared four-momentum transfer of the virtual photon (or $Z$ boson), with $Q^2=s$.

Within the perturbative QCD (pQCD) up to $O(\alpha_s^1)$, the total hadronic cross
section $\sigma_{tot}$ is given by
\begin{equation}
    \sigma_{tot} = \sum_q \sigma_0^q(s)\Bigg[ \Big(1 + \frac{\alpha_s(Q)}{\pi}\Big)  \Bigg]\ ,
\end{equation}
where the electroweak cross sections $\sigma_0^q$ for producing a $q\bar{q}$ pair at the $O(\alpha_s^0)$ order of pQCD are given in Ref.~\cite{Nason:1993xx}.

The structure function can be further divided into the sum of the longitudinal and transversal part,
\begin{equation}
    F^{h^{\pm}}(z,Q) = F_T^{h^{\pm}}(z,Q) + F_L^{h^{\pm}}(z,Q)\ ,
\end{equation}
where $F_{P=T,L}^{h^{\pm}} $ defined as,
\begin{equation} \label{TL function}
    F_{P=T,L}^{h^{\pm}}(z,Q) = \frac{1}{\sigma_{tot}} \sum_i  C_{P=T,L}^i(z,Q) \otimes D_i^{h^{\pm}}(z,Q)\ .
\end{equation}

The convolution symbol $\otimes$ in equations above is defined as,
\begin{equation} \label{otimes}
    f(z) \otimes g(z) = \int_z^1 \frac{dx}{x}f(x)g(\frac{z}{x})\ .
\end{equation}

The coefficient functions in Eq. (\ref{TL function}) have been
calculated in pQCD up to $O(\alpha_s^1)$ at the next-to-leading order (NLO)~\cite{Nason:1993xx,Altarelli:1979kv,Furmanski:1981cw} and  to $O(\alpha_s^2)$ at the next-to-next-to leading order (NNLO) accuracy~\cite{Rijken:1996ns,Mitov:2006wy}. We only consider the NLO accuracy in our analysis.

In following analysis we used the Zero-Mass Variable-Flavor-Number Scheme (ZM-VFNS) which considers all active flavors massless. However, for heavy quarks we need to determine the number of active flavors based on the threshold. In this work, the charm and bottom masses are considered to be fixed at $m_c$ = 1.4 GeV and $m_b$ = 4.5 Gev, respectively. In addition, we do not consider the top quarks which decay before hadronization. We choose $\alpha_s(M_Z) = 0.118$ at the $Z$ boson mass $M_Z = 91.2$ GeV as a reference value which is close to the world-average of the
Particle Data Group \cite{ParticleDataGroup:2018ovx}.

The scale dependence of the fragmentation functions is described by the DGLAP evolution equation, which in the integral-differential form is given by \cite{Gribov:1972ri,Gribov:1972rt,Dokshitzer:1977sg,Altarelli:1977zs},
\begin{equation}
    {\partial D_i^{h^{\pm}}(z,Q) \over \partial \ln{Q^2}} = {\alpha_s(Q) \over 2\pi} \sum_j P_{ji}(z,\alpha_s) \otimes D_j^{h^{\pm}}(z,Q),
     \label{DGLAP eq}
\end{equation}
where $P_{ji}(z,\alpha_s)$ are time-like splitting functions that describe the splitting process $i \to j + X$ with the indices $i,j$ running over all parton flavors (quarks, antiquarks and the gluon). This system represents a set of coupled $2n_f +1$ equations.

To decouple the evolution equations, we decompose the quark FFs into the singlet component ($D_{\Sigma}$) and a set of valence non-singlet components ($D_{NS,V}$) based on the $SU(n_f)$ flavor symmetry~\cite{Furmanski:1981cw,Kovchegov:2012mbw,Curci:1980uw}:
\begin{align} \label{singlet}
    D_{\Sigma}^{h^{\pm}} =& \sum_{q}^{n_f} (D_q^{h^{\pm}} + D_{\bar{q}}^{h^{\pm}}) = \sum_q^{n_f} D_{q^{+}}^{h^{\pm}},\\
\label{NS_V}    D_{NS,V}^{h^{\pm}} =& D_q^{h^{\pm}} - D_{\bar{q}}^{h^{\pm}} =  D_{q^{-}}^{h^{\pm}},
\end{align}
where $D_{q^{\pm}}^{h^{\pm}} \equiv D_q^{h^{\pm}} \pm D_{\bar{q}}^{h^{\pm}}$ and the index $q$ runs over all active quark flavors ($q = {u,d,s,\ldots}$).

In addition to the valence non-singlet $D_{NS,V}^{h^{\pm}}$, the remaining non-singlet combinations are constructed to be orthogonal to the singlet. Following the standard convention, we define the independent non-singlet combinations $T_k$ as:
\begin{align}
    \label{SFFs3}        T_3 &= D_{u^+}^{h^{\pm}} - D_{d^+}^{h^{\pm}}\ ;\\
    \label{SFFs4}        T_8 &= D_{u^+}^{h^{\pm}} + D_{d^+}^{h^{\pm}} -2D_{s^+}^{h^{\pm}}\ ;\\
    \label{SFFs5}        T_{15} &= D_{u^+}^{h^{\pm}} + D_{d^+}^{h^{\pm}} + D_{s^+}^{h^{\pm}} -3D_{c^+}^{h^{\pm}}\ ;\\
    \label{SFFs6}        T_{24} &= D_{u^+}^{h^{\pm}} + D_{d^+}^{h^{\pm}} + D_{s^+}^{h^{\pm}} + D_{c^+}^{h^{\pm}} -4D_{b^+}^{h^{\pm}}\ ;\\
    \label{SFFs7}        T_{35} &= D_{u^+}^{h^{\pm}} + D_{d^+}^{h^{\pm}} + D_{s^+}^{h^{\pm}} + D_{c^+}^{h^{\pm}} + D_{b^+}^{h^{\pm}} -5D_{t^+}^{h^{\pm}}\ .
\end{align}
 The labels $3,8,15,\ldots$ follow the conventional numbering of the diagonal (Cartan) generators in a generalized Gell-Mann basis of $SU(n_f)$. For a given number of active flavors $n_f$, one should keep only the $(n_f-1)$ independent diagonal non-singlet combinations, conventionally denoted by $T_3, T_8, \ldots, T_{n_f^2-1}$. Including combinations associated with a larger flavor group would render the basis over-complete (linearly dependent).

In this evolution basis, the equations partially decouple. The singlet FF $D_{\Sigma}$ mixes with the gluon FF $D_g$:
\begin{align} \label{S}
    \frac{\partial}{\partial\ln{Q^2}}&\left (
 \begin{matrix}
   D_{\Sigma}^{h^{\pm}}(z,Q) \\
   D_g^{h^{\pm}} (z,Q)\\
  \end{matrix}
  \right  ) = \frac{\alpha_s(Q)}{2\pi}  \\ 
  &\times \left( 
  \begin{matrix}
  P_{qq}(z,\alpha_s) & 2n_fP_{gq}(z,\alpha_s)\\
  P_{qg }(z,\alpha_s) & P_{gg}(z,\alpha_s)
  \end{matrix}
  \right) \otimes \left (
 \begin{matrix}
   D_{\Sigma}^{h^{\pm}}(z,Q) \\
   D_g^{h^{\pm}}(z,Q) \\
  \end{matrix}
  \right  )\  \notag,
\end{align}
The non-singlet components evolve independently. Due to charge conjugation symmetry, the splitting functions satisfy $P_{qq} = P_{\bar{q}\bar{q}}$ and $P_{q\bar{q}} = P_{\bar{q}q}$. This allows us to define two independent non-singlet evolution kernels:
\begin{equation}
     \frac{\partial}{\partial\ln{Q^2}}D_{NS^{\pm}}^{h^{\pm}}(z,Q) =\frac{\alpha_s(Q)}{2\pi} P_{NS^{\pm}} \otimes D_{NS^{\pm}}^{h^{\pm}}(z,Q)\ ,
     \label{NS}
\end{equation}
where $P_{NS^{\pm}} \equiv P_{qq} \pm P_{q\bar{q}}$. Here, the notation $D_{NS^+}^{h^{\pm}}$ represents the set of charge conjugation-even combinations $\{T_k\}$ (defined in Eqs.~\eqref{SFFs3}--\eqref{SFFs7}), which evolve with $P_{NS^+}$, while $D_{NS^-}^{h^{\pm}}$ corresponds to the charge conjugation-odd valence combination $D_{NS,V}^{h^{\pm}}$ (defined in Eq.~\eqref{NS_V}), which evolves with $P_{NS^-}$. Note that $P_{q\bar{q}}$ vanishes at the leading order but contributes at higher orders.

We can perform a Taylor expansion of the splitting functions in Eq.~\eqref{S} and~\eqref{NS} in terms of the strong coupling constants,
\begin{align}
\label{taile1}       P_{ij}(z,\alpha_s) &=  P_{ij}^{(0)}(z) + {\alpha_s(Q) \over 2\pi}P_{ij}^{(1)}(z) + \mathcal{O}(\alpha_s^2) ,\\
\label{taile2}       P_{NS^{\pm}}(z,\alpha_s) &= P_{NS^{\pm}}^{(0)}(z) + {\alpha_s(Q) \over 2\pi} P_{NS^{\pm}}^{(1)}(z) + \mathcal{O}(\alpha_s^2).
\end{align}
 
The LO and NLO time-like splitting functions have been computed in Refs.~\cite{Curci:1980uw,Gluck:1989ze,Gluck:1992zx,Floratos:1981hs}. Eq.~\eqref{S} and~\eqref{NS} are differential-integral equations, and the most straightforward approach to solve the equation is to use numerical approximation techniques directly in $z$-space~\cite{Hirai:2011si}. However, as discussed in Ref.~\cite{Sato:2016tuz}, solving the equations for the Mellin moments will be more efficient because the Mellin transformation convert the convolution in $z$-space into a product of the Mellin moments. The Mellin transform and inverse Mellin transform are defined as, respectively~\cite{Ellis:1996mzs},
\begin{align} \label{Mellin transform}
    f(N) &= \int_0^1 dz z^{N-1} f(z)\ ,\\
    f(z) &= \frac{1}{2\pi i}\int_{c-i\infty}^{c+i\infty} dN z^{-N} f(N)\ .
\end{align}
The integration contour in the inverse Mellin transform is the standard Bromwich contour, i.e., a straight line parallel to the imaginary axis, defined by $N = c+ it$ with $t \in (-\infty,\infty)$. The real constant $c$ is chosen to be to the right of the rightmost singularity of $f(N)$, ensuring that all poles and branch cuts lie to the left of the contour. For numerical evaluation, one can maximize the computational efficiency by deforming the contour into a tilted path in the complex plane. As derived in Ref. \cite{Vogt:2004ns}, this deformation allows the integral to be expressed as an integral over a real variable, significantly improving convergence by damping the oscillatory behavior.

Using Eq.~\eqref{Mellin transform}, the convolution in Eq.~\eqref{otimes} can be turned into the product of the Mellin moments~\cite{Vogt:2004ns},
\begin{equation}
     M[f(z) \otimes g(z)] = f(N)\ g(N)\ ,
 \end{equation}
 where $f(N)$, $g(N)$ are the Mellin transforms of $f(z),g(z)$, respectively.

 Using the Mellin transformation,  Eqs. (\ref{dx cs}), (\ref{S}) and (\ref{NS}) in terms of Mellin moments become,
 \begin{equation}
 F^{h^{\pm}}(N,Q) = \frac{1}{\sigma_{tot}} \sum_i C_i(N,Q)D_i^{h^{\pm}}(N,Q)\ , 
  \label{Fh_N}
 \end{equation}
 \begin{align}
\frac{\partial}{\partial\ln{Q^2}}&\left (
 \begin{matrix} \label{N_S}
   D_{\Sigma}^{h^{\pm}}(N,Q) \\
   D_g^{h^{\pm}}(N,Q) \\
  \end{matrix}
  \right  ) =\frac{\alpha_s(Q)}{2\pi}   \\
  &\hspace{-1cm}\times \left( 
  \begin{matrix}
  M_{qq}(N,\alpha_s) & 2n_fM_{gq}(N,\alpha_s)\\
  M_{qg}(N,\alpha_s) & M_{gg}(N,\alpha_s)
  \end{matrix}
  \right)
  \left (
 \begin{matrix}
   D_{\Sigma}^{h^{\pm}}(N,Q) \\
   D_g^{h^{\pm}}(N,Q)\\
  \end{matrix}
  \right  )\  \notag,
\end{align}

\begin{equation}
     \frac{\partial}{\partial\ln{Q^2}}D_{NS}^{h^{\pm}}(N,Q) =\frac{\alpha_s(Q)}{2\pi} M_{NS^{\pm}}(N,\alpha_s)  D_{NS}^{h^{\pm}}(N,Q)\ .
     \label{N_NS}
\end{equation}

The Mellin moments of the coefficient functions $C_i(N,Q)$ at NLO are listed in Refs. \cite{Nason:1993xx,Altarelli:1979kv,Furmanski:1981cw} and the Mellin moments of the  splitting functions at LO and NLO can be found in the Refs. \cite{Owens:1978qz,Uematsu:1978yw,Georgi:1977mg,Curci:1980uw,Gluck:1989ze,Gluck:1992zx,Floratos:1981hs}. For consistency in our analysis,  the coefficient functions and splitting functions are both computed at the NLO accuracy. In addition, we consider the charge conjugation $D_{q (\bar{q})}^{h^+} = D_{\bar{q} (q)}^{h^-} $, which separates the quarks and antiquarks, and the relationship \cite{Soleymaninia:2018uiv}
\begin{equation}
    D_q^{h^{\pm}} = D_{\bar{q}}^{h^{\pm}} = \frac{D_{q^+}^{h^{\pm}}}{2}\ .
\end{equation}

Since quark and antiquark FFs are equal, we can ignore the $M_{NS^-}$ function \cite{Sato:2016wqj}. Moreover, our analysis chooses the fragmentation scale equal to the center of mass energy, i.e., $\mu_f = \sqrt{s} = Q$.

\section{Fragmentation functions via Physics-Informed Neural Networks}
\label{THREE}

\subsection{Physics Informed Neural Network}
\begin{figure*}[htbp]
    \centering
    \includegraphics[width=0.9\textwidth]{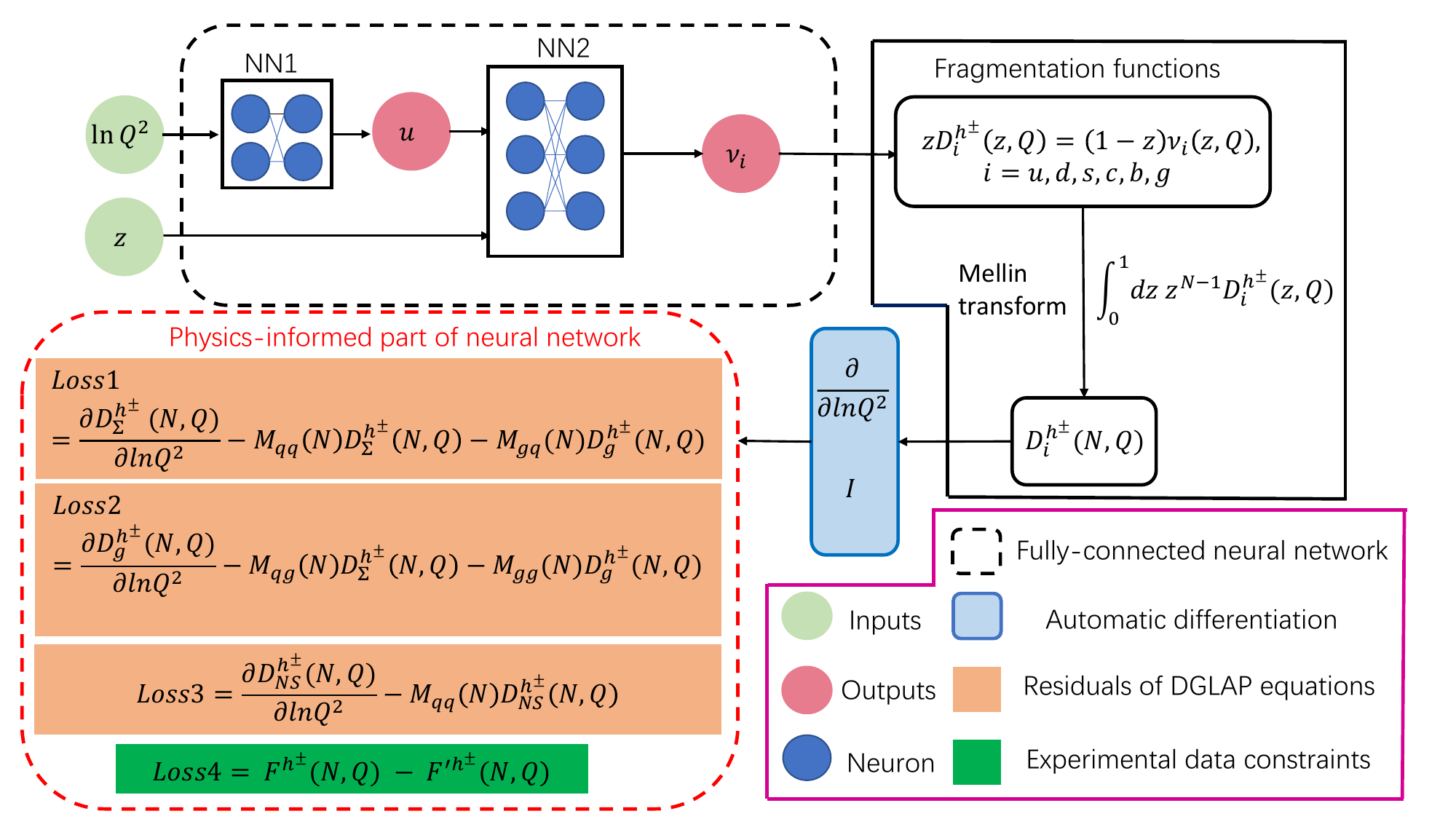}
    \caption{The neural network framework divided into four modules for FFs extractions. The first module is the black dashed box representing the fully connected neural network. The second module is the black solid box representing the construction of the FFs and the Mellin transformation of the FFs. The third module is the blue solid box representing the auto-differentiation technique in deep learning, which can be used for differential computation in the DGLAP evolution equations. The fourth module is the red dashed box, representing the Physics-informed part of neural network. The primary purpose of this module is to allow the neural network to satisfy the DGLAP evolution equations and to compute the cross section of the SIA process in the Mellin space.}
    \label{PINN}
\end{figure*}


In this paper, we adopt a Physics-Informed Neural Network(PINN) \cite{RAISSI2019686,9429985,DBLP:journals/corr/abs-2201-05624} as a new technique to extract the FFs numerically from the experimental data on single inclusive hadron spectra in SIA.

Compared to the traditional application of deeply convoluted neural networks to classification and regression tasks, PINN integrates the neural network as a component within a framework designed to solve partial differential or ordinary differential equations. In our proposed framework, the neural network we constructed needs to satisfy the following characteristics. First, it can compute the Mellin moments of the experimental observables of the SIA, specifically denoted as $F^{h^{\pm}}(N, Q)$. Secondly, the neural network must enforce the DGLAP evolution equations on FFs. Lastly, as the DGLAP evolution equations is enforced through Mellin moments, the neural network should also have the capability to perform the inverse Mellin transform. 

We divide this PINN framework for FFs extractions into four modules and illustrate it in Fig. \ref{PINN}. The first module (black dashed box) adopts a cascaded design with two fully connected networks, NN1 and NN2, each with 8 hidden layers and 64 neurons per layer. NN1 takes the evolution variable $\ln(Q^2)$ as input and, with Softplus activations, outputs a latent feature $u(\ln Q^2)$ that encodes the $Q^2$-dependence. We then concatenate $u(\ln Q^2)$ with the kinematic variable $z$ and feed the combined input into NN2 to output an auxiliary function $v_i$ from $(z, u(\ln Q^2))$. NN2 uses Softplus in its hidden layers, while its outputs are mapped through $\exp(-x)$ to enforce $v_i > 0$, ensuring the positivity of $zD_i$ over the entire kinematic domain. This factorized architecture separates the slowly varying perturbative evolution in $\ln(Q^2)$ from the strongly nonlinear $z$-dependence. By reducing the coupling between variables with disparate scales and sensitivities, it helps mitigate gradient imbalance that can arise in a single flat Feed-Forward Network (FFN), which can improve optimization stability.

The second module is inside the black solid box representing the form of the FFs constructed and their Mellin transform. It is known that the FFs are zero at $z = 1$, i.e., $D_i^{h^{\pm}}(z=1) = 0$. In solving partial differential equations with neural networks, there are typically two approaches to handling boundary conditions: one is to take the boundary conditions as one of the training objectives and the other is to construct a neural network that naturally satisfies the boundary conditions. When the network has more training objectives, the former approach often results in poor training outcomes and slow convergence, whereas the latter ensure strict adherence to boundary conditions and faster converge as shown in our test results. To enforce this boundary condition in our framework, the FFs for every parton flavor are defined as, 
\begin{equation} \label{FFs_NN}
    zD_i^{h^{\pm}}(z,Q) = (1-z)\nu_i(z,Q) \quad (i = u,d,s,c,b,g)\ ,
\end{equation}
where $\nu_i(z, Q)$ is the output of the neural network in the first module for parton flavor $i$. Furthermore, we have noted that the FFs for light flavors or at fixed large scales are monotonically decreasing in $z$. In our study here, the data we use start at $Q^2 = 14~\mathrm{GeV}^2$. Our PINN parameterizes the distributions directly in the analyzed kinematic region, not at the initial scale. At these scales, charm is already substantially smoothed by DGLAP evolution, so a monotonic shape provides a reasonable effective description. For bottom, while a peak may exist at lower scales or at larger $z$, the current experimental data do not tightly constrain the detailed large-$z$ shape. We therefore impose the monotonicity constraint mainly as a mild regularizer to stabilize the fit and suppress nonphysical oscillations in sparsely constrained regions, with minimal impact on the data-driven $z$ range. This is achieved by adjusting the positive and negative values of network parameters.
Additionally, since we solve the DGLAP evolution equations in terms of the Mellin moments $D_i^{h^{\pm}}(N, Q)$, we need to perform the Mellin transformation of the FFs using the Guass-Legendre method.

\begin{figure*}[htbp]
    \centering
    \includegraphics[width=0.9\textwidth]{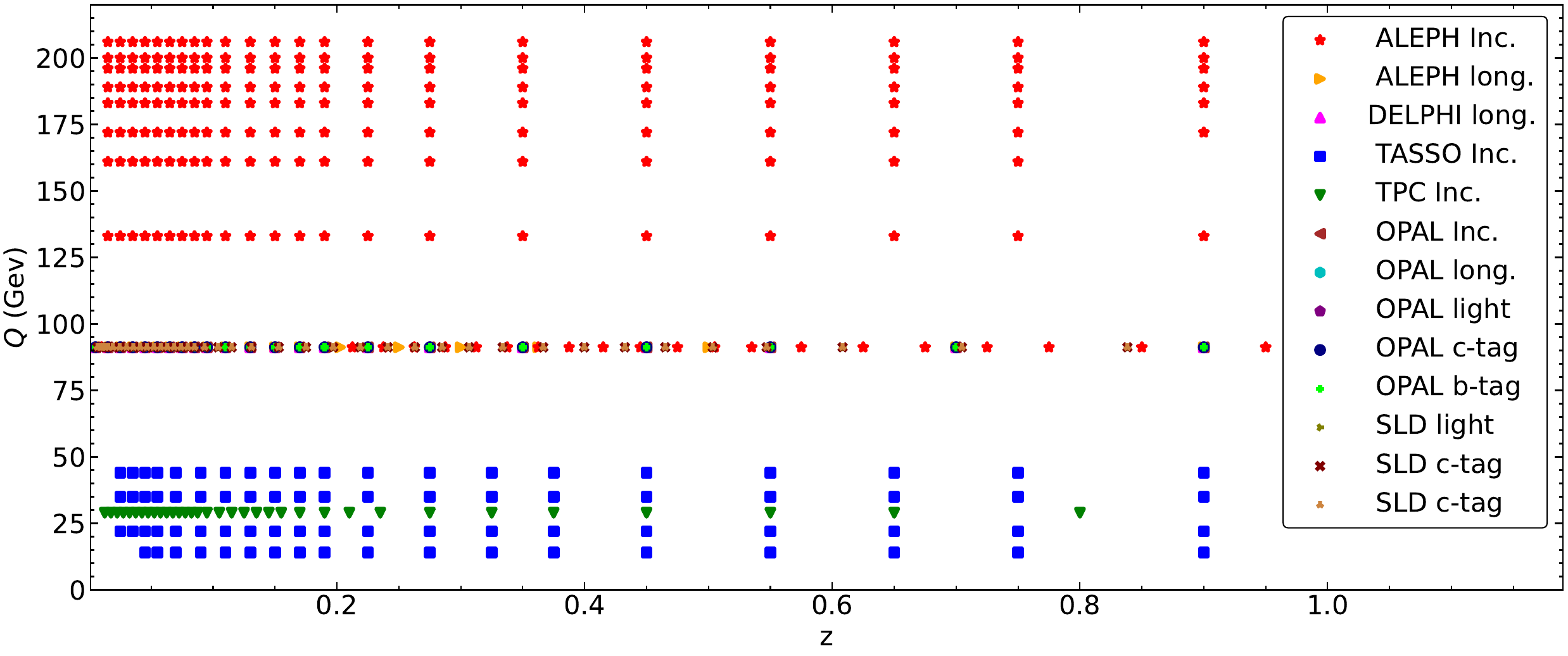}
    \caption{Kinematic ranges in $(z, Q)$ in experimental SIA data used to determine the FFs.}
    \label{data}
\end{figure*} 

The third module of our PINN framework is the blue solid box in Fig. \ref{PINN} representing the auto-differentiation for the DGLAP evolution equations in deep learning \cite{JMLR:v18:17-468}. Unlike classical numerical and symbolic differentiations, auto-differentiation employs the chain rule and back-propagation to compute the derivatives of the output with respect to the input or the trainable parameters within the network. This method achieves analytical precision in its results. 

The primary function of the fourth module in the PINN as represented by the red dashed box is to ensure that the neural network enforces the DGLAP evolution equations and to compute the cross section for the SIA process in terms of the  Mellin moments. The three orange boxes inside this module correspond to the residuals of the DGLAP evolution equations, related to Eqs. (\ref{N_S}) and (\ref{N_NS}). The green box represents the difference between the cross section computed by the neural network ($F^{h^{\pm}}(N, Q)$) and the experimental data ($F^{\prime h^{\pm}}(N, Q)$), as quantified by Eq. (\ref{Fh_N}).

Our objective is to obtain a solution to the DGLAP evolution equations that is simultaneously consistent with the experimental data. This is achieved by minimizing a composite loss function, $\mathcal{L}_{\text{total}} = \omega_1 \mathcal{L}_1 + \omega_2 \mathcal{L}_2 + \omega_3 \mathcal{L}_3 + \omega_4 \mathcal{L}_4$, where the first three terms enforce the physical constraints and the fourth term quantifies the deviation from experimental measurements. To address the disparity in orders of magnitude across these terms, the weighting coefficients are calibrated as $(\omega_1, \omega_2, \omega_3, \omega_4) = (5000, 5000, 100, 3000)$. This configuration effectively equilibrates the relative contributions of the physics residuals and data mismatch during backpropagation, thereby preventing any single component from dominating the gradient updates. The network parameters are optimized by minimizing $\mathcal{L}_{\text{total}}$ using gradient-based algorithms such as Adam~\cite{bottou_1999, kingma2017adam}.

%

\begin{figure*}[htbp]
    \centering
    \includegraphics[width=0.85\linewidth]{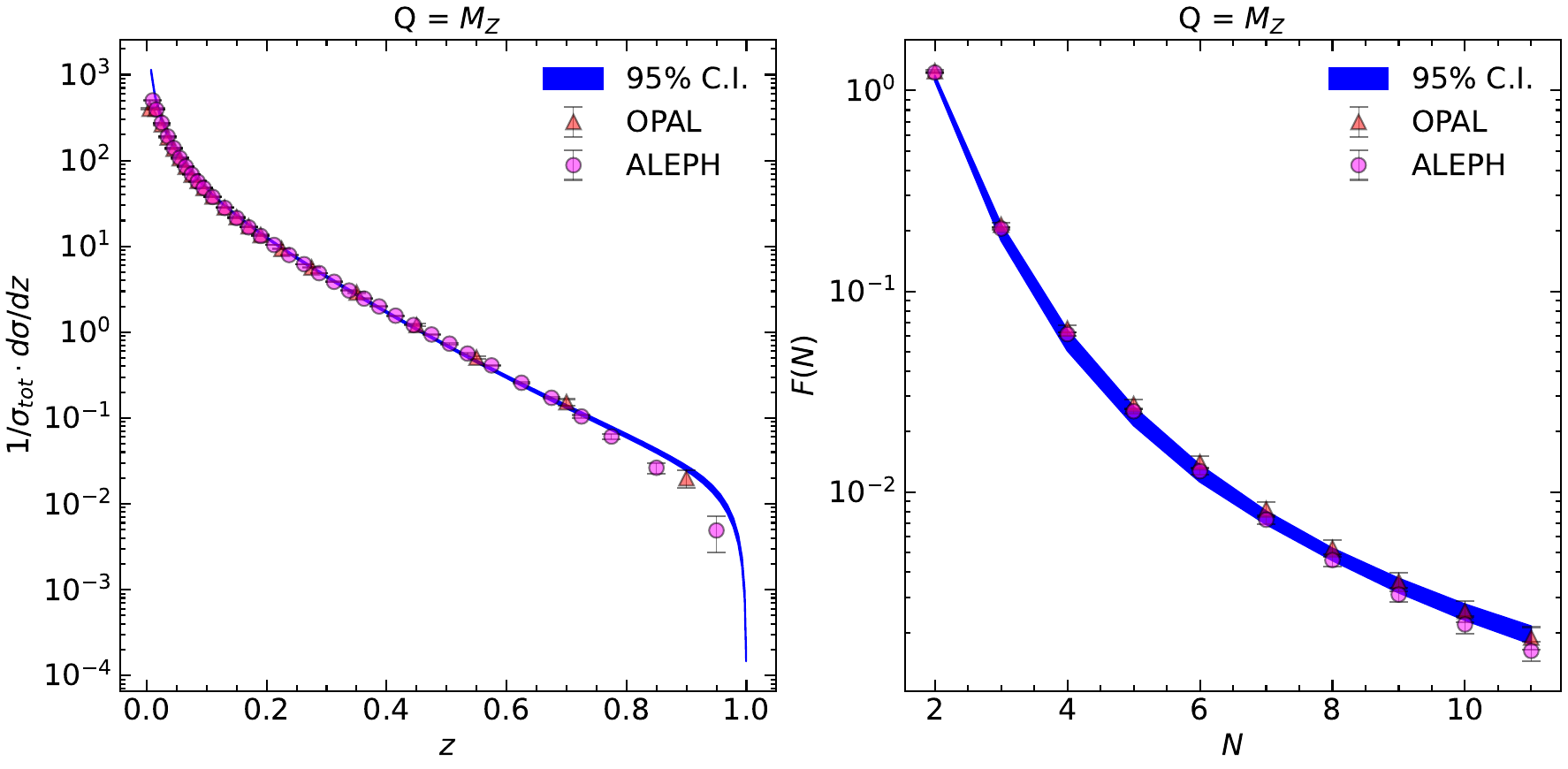}
    \caption{Comparison of the training results of the NN with the experimental data in FFs as function of $z$ and the Mellin moments, respectively. The uncertainty of the results is calculated from the 100 samples with a confidence interval of 2$\sigma$.}
    \label{Fx and FN}
\end{figure*}

\subsection{Experimental Data Sets and Uncertainty Quantification}
In this analysis, we utilize a comprehensive set of experimental data on single-inclusive unidentified charged hadron production in electron-positron annihilation from the ALEPH~\cite{ALEPH:1995njx,ALEPH:2003obs}, the TASSO~\cite{TASSO:1990cdg}, the TPC~\cite{TPCTwoGamma:1988yjh}, and the OPAL~\cite{OPAL:1998arz} Collaborations.
The kinematic coverage of the analyzed SIA dataset in the $(z, Q)$ plane is illustrated in Fig.~\ref{data}. The center-of-mass energy $Q$ ranges from 14 GeV to 206 GeV, providing sufficient coverage for the extracted FFs across the energy intervals relevant to current experiments. Given that the minimum scale of the selected data ($Q = 14$ GeV) significantly exceeds the bottom quark mass threshold, we adopt a fixed number of active quark flavors, $n_f=5$, throughout the analysis. 

To achieve a comprehensive extraction, we incorporate not only inclusive measurements but also longitudinal inclusive and 
flavor-tagged cross section from the ALEPH~\cite{ALEPH:1995njx}, OPAL~\cite{OPAL:1998arz}, DELPHI~\cite{DELPHI:1998cgx}, and SLD~\cite{SLD:2003ogn} Collaborations, as shown in Fig.~\ref{data}. The inclusion of longitudinal inclusive cross section is particularly crucial, as it exhibits enhanced sensitivity to the gluon fragmentation function via next-to-leading order corrections, thereby providing stronger constraints than total cross section data alone. Furthermore, the flavor-tagged measurements—covering the sum of light quarks ($u, d, s$) and individual heavy quarks ($c, b$)—are indispensable for the flavor decomposition of the fragmentation functions. This combination of datasets allows for an unambiguous separation of light and heavy flavor contributions, ensuring a level of precision that is not attainable with inclusive data alone.

To propagate experimental uncertainties into our final extracted FFs, we employ the Monte Carlo replica method. We generate an ensemble of pseudodata sets by performing Gaussian sampling around the central experimental values. For each data point $i$, the pseudodata value $\widetilde{D}_i$ is generated according to:
\begin{equation}
    \widetilde{D}_i = D_i + R_i\sigma_i\ ,
\end{equation}
where $D_i$ is the experimental central value, $\sigma_i$ represents the total experimental uncertainty (combining statistical and systematic errors in quadrature), and $R_i$ is a random number drawn from a standard normal distribution $\mathcal{N}(0, 1)$. Each instance of the neural network is then trained on a unique set of pseudodata replicas, thereby propagating the experimental errors into our results.

Finally, 
our theoretical formalism is based on fixed-order perturbative QCD and does not incorporate small-$z$ logarithmic resummation or higher-twist corrections. Given that the logarithmic contributions become particularly significant in the small-$z$ regime, we explicitly exclude the region $z < 0.01$ from our analysis to ensure the validity of the fixed-order calculation and maintain consistency between theoretical predictions and data.

\subsection{Training results for the datasets}

Figure~\ref{Fx and FN} presents the training outcomes of the neural network compared against the experimental data at the center-of-mass energy of $M_Z$. 
for both the FFs in the fractional momentum $z$ (left panel) and the conjugate Mellin moments (right panel).

Our extraction strategy is formulated primarily for the Mellin moments of FFs. In our approach, the experimental measurements of the $z$-distribution are transformed into the Mellin moments via numerical integration. These transformed moments serve as the direct training targets for the neural network, allowing the optimization to be performed entirely on the Mellin moments. To quantify the statistical uncertainties, we employ the Monte Carlo replica method. One hundred pseudo-data replicas were generated by sampling from Gaussian distributions defined by the central values and their experimental standard deviations. The neural network was trained on each replica, and the resulting ensemble yields the 2-$\sigma$ confidence interval shown as the blue band in the
right panel.

The validation of the training performance is demonstrated in the left panel of Fig.~\ref{Fx and FN} where the trained network predictions of the Mellin moments were mapped back to the FFs as a function of $z$ via the Inverse Mellin Transform. The resulting band in the left panel exhibits excellent agreement with the original experimental data. This consistency confirms that the neural network has successfully captured the underlying distribution in the Mellin moments and that the extracted FFs can be reliably obtained from transform of the Mellin moments for phenomenological applications.

\section{Results and Discussion}
\label{FOUR}
\subsection{DGLAP Verification}
In this section, the numerical results from the PINN are verified against the DGLAP evolution equations. The Mellin moments of the singlet and gluon fragmentation functions at $Q=M_Z$, extracted by the neural network, are used as initial conditions for the DGLAP evolution. These moments are evolved to 14 GeV and 1 TeV using the standard fourth-order Runge--Kutta (RK4) method. To ensure the precision of this numerical baseline, a step-doubling convergence test was performed by halving the evolution step size. The resulting relative error is found to be of the order $\mathcal{O}(10^{-4})$, confirming that numerical integration errors are negligible. The evolved results were subsequently compared with the fragmentation functions (Mellin moments) provided by the neural network at the scales of 14 GeV and 1 TeV.

\begin{figure*}[htbp]
	\centering
	\begin{minipage}{0.49\linewidth}
		\centering
		\includegraphics[width=1\linewidth]{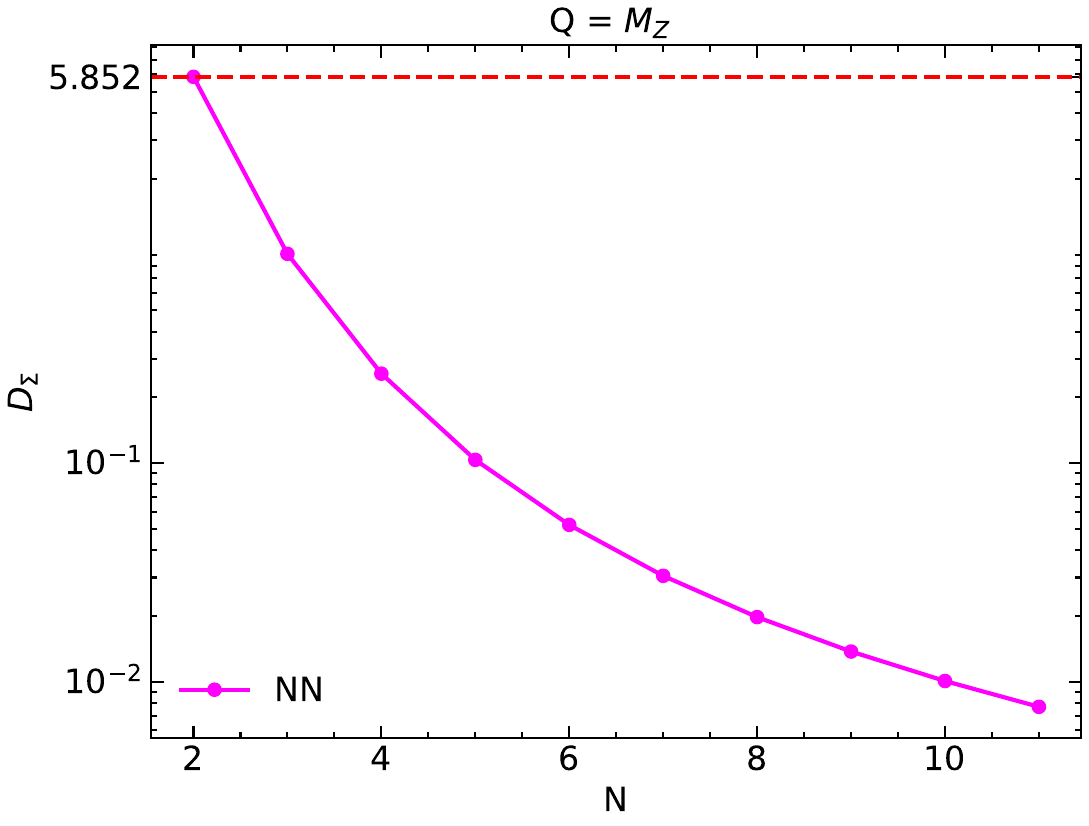}
	\end{minipage}
	\begin{minipage}{0.49\linewidth}
		\centering
		\includegraphics[width=1\linewidth]{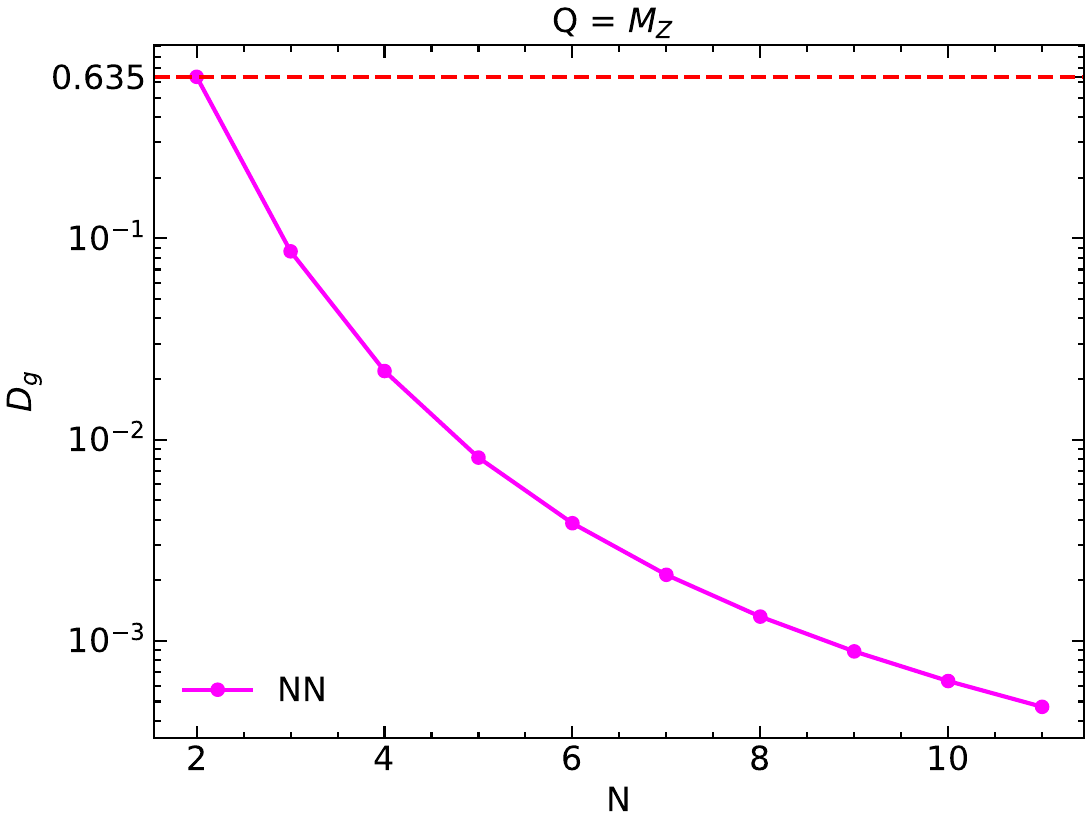}
	\end{minipage}
 \caption{The Mellin moments of the singlet quark ($D_{\Sigma}^{h^{\pm}}$) and gluon ($D_g^{h^{\pm}}$) fragmentation functions at the scale $Q = M_Z$, obtained from the PINN. The dashed straight lines indicate the momentum sum ($N=2$ Mellin moments).}
 \label{baseline_91.2}
\end{figure*}

Fig.~\ref{baseline_91.2} shows the Mellin moments of the singlet quark and gluon fragmentation function given by the neural network at scale $Q = M_Z$, which we use as initial values for the DGLAP evolution equations. We can verify that our FFs satisfies the momentum sum rule,
\begin{equation}
    \sum_h \int_0^1 dz\ z\ D_i^h(z,Q) = 1\ , \text{$i = q,\bar{q},g$}
    \label{sum rule}
\end{equation}
for any value of $Q$, which ensures that the total momentum carried by all hadrons produced in the fragmentation of a given parton $i$ is the same as that carried by the parton itself. The DGLAP evolution guarantees the momentum sum rule as a direct consequence of energy conservation.

By the definition of Mellin moments in Eq. (\ref{Mellin transform}), we can see that Eq. (\ref{sum rule}) is just the value of $N=2$ of the Mellin moments, that is \cite{Hirai:2007cx}
\begin{equation}
    \sum_h D_i^h(N=2,Q) = \sum_h \int_0^1 dz\ z\ D_i^h(z,Q) = 1\ .
\end{equation}

In our analysis, we only consider the fragmentation functions of parton $i$ into unidentified charged hadrons which are only a subset of all hadrons. Therefore, we should have $\sum_h D_i^{h^{\pm}}(N=2,Q) < 1$. From Fig. \ref{baseline_91.2}, we can see that $D_{\Sigma}^{h^{\pm}}(N = 2, Q \text{GeV}) = 5.852 < 2n_f$ (five flavors $n_f=5$ ), $D_g^{h^{\pm}}(2,Q \text{GeV}) = 0.635 < 1$, both satisfying the momentum sum rule. 

In principle, we could use Eq.~(\ref{sum rule}) to constrain the FFs, especially in small-$z$ regions where no experimental information is available. This is, however, not implemented in our calculations mainly because it requires the knowledge of the FFs for all the hadronic species $h$, while we consider only a subset of them. It is worth noting that although we did not add the momentum sum rule as a physical constraint to the NN, the FFs given by the NN still satisfy the momentum sum rule which is embedded in the experimental data used for the NN training.

\begin{figure*}[htbp]
	\centering
	\begin{minipage}{0.49\linewidth}
		\centering
		\includegraphics[width=1\linewidth]{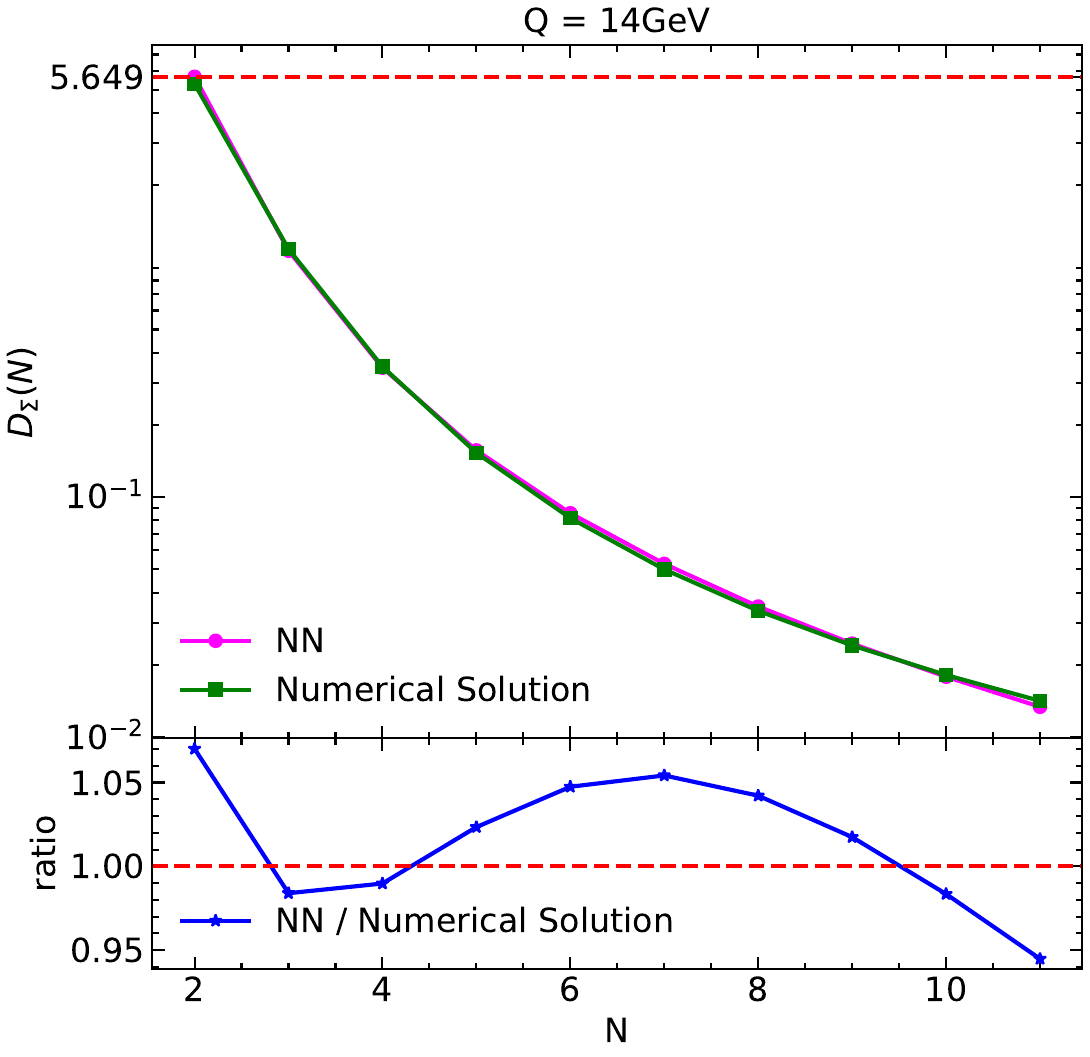}
	\end{minipage}
	\begin{minipage}{0.49\linewidth}
		\centering
		\includegraphics[width=1\linewidth]{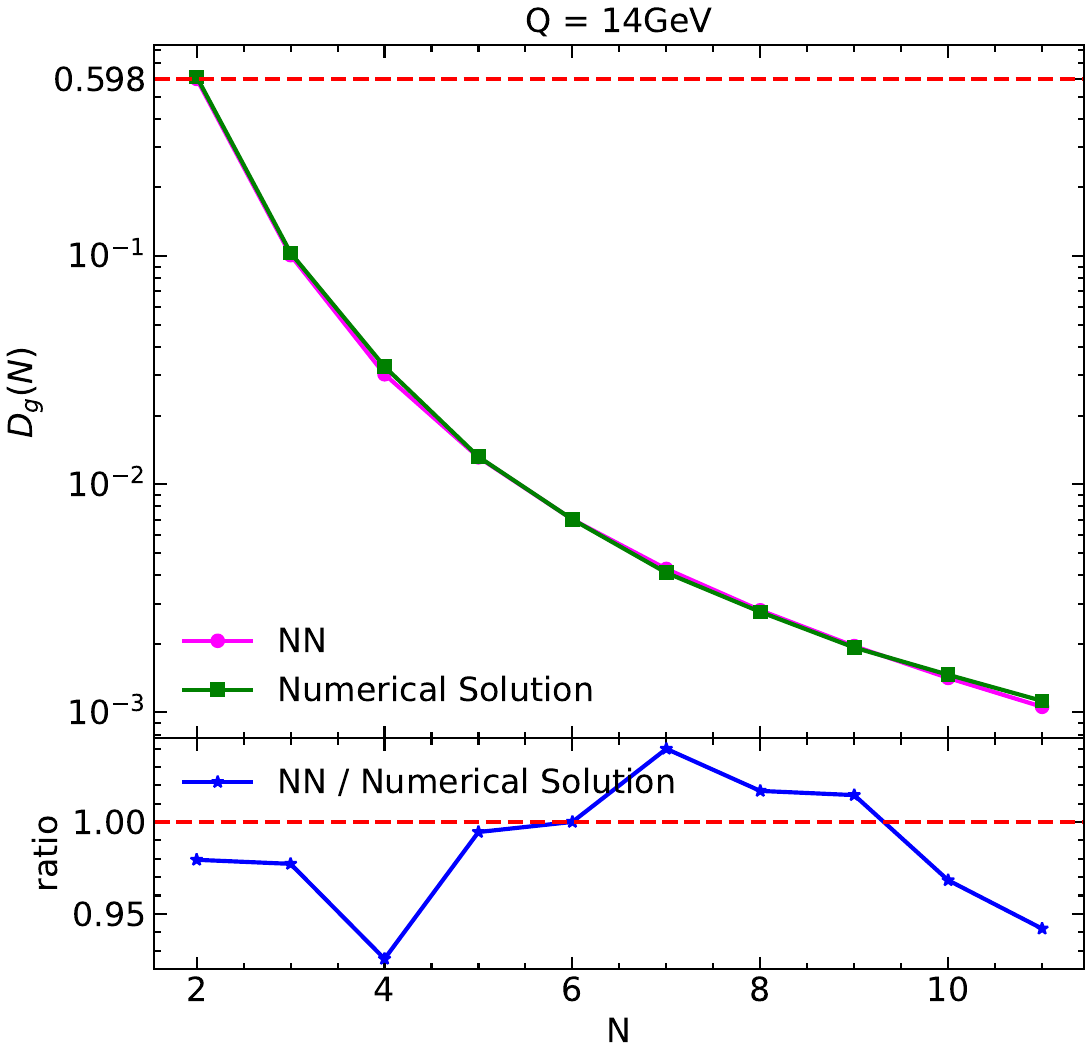}
	\end{minipage}
 \caption{Upper panel: Comparison of the Mellin moments of the FFs at $Q = 14$ GeV. The green lines represent the moments evolved from the initial conditions in Fig. \ref{baseline_91.2} using the numerical DGLAP solver, while the magenta lines correspond to the direct PINN predictions at $Q = 14$ GeV. The dashed straight lines indicate the momentum sum ($N=2$). Lower panel: The ratios of the PINN predictions to the numerical solutions for moments of the singlet quark and gluon FFs.}
      \label{evolution_14}
\end{figure*}

The upper panel of Fig. \ref{evolution_14} shows the results of  the quark singlet and gluon fragmentation function evaluated at scale $Q = $ 14 GeV. The green line is the result obtained by solving DGLAP evolution equations with numerical approximation techniques using FFs at scale $Q = M_Z$ in Fig. \ref{baseline_91.2} as the initial condition. The magenta line is the result given directly by the NN. The lower panel of Fig.~\ref{evolution_14} shows the ratio between the results from NN and numerical evolution. We can see
that the results obtained by the NN are consistent with those given by direct numerical evolution within 5\% accuracy.
In addition, we can see again that the FFs satisfies the momentum sum rule, i.e., $D_{\Sigma}^{h^{\pm}}(N = 2,\ Q = 14\ \text{GeV}) = 5.649 < 2n_f = 10$ (five flavors $n_f=5)$), $D_g^{h^{\pm}}(2,\ Q = 14\ \text{GeV}) = 0.598 < 1$.

\begin{figure*}[htbp]
	\centering
	\begin{minipage}{0.49\linewidth}
		\centering
		\includegraphics[width=1\linewidth]{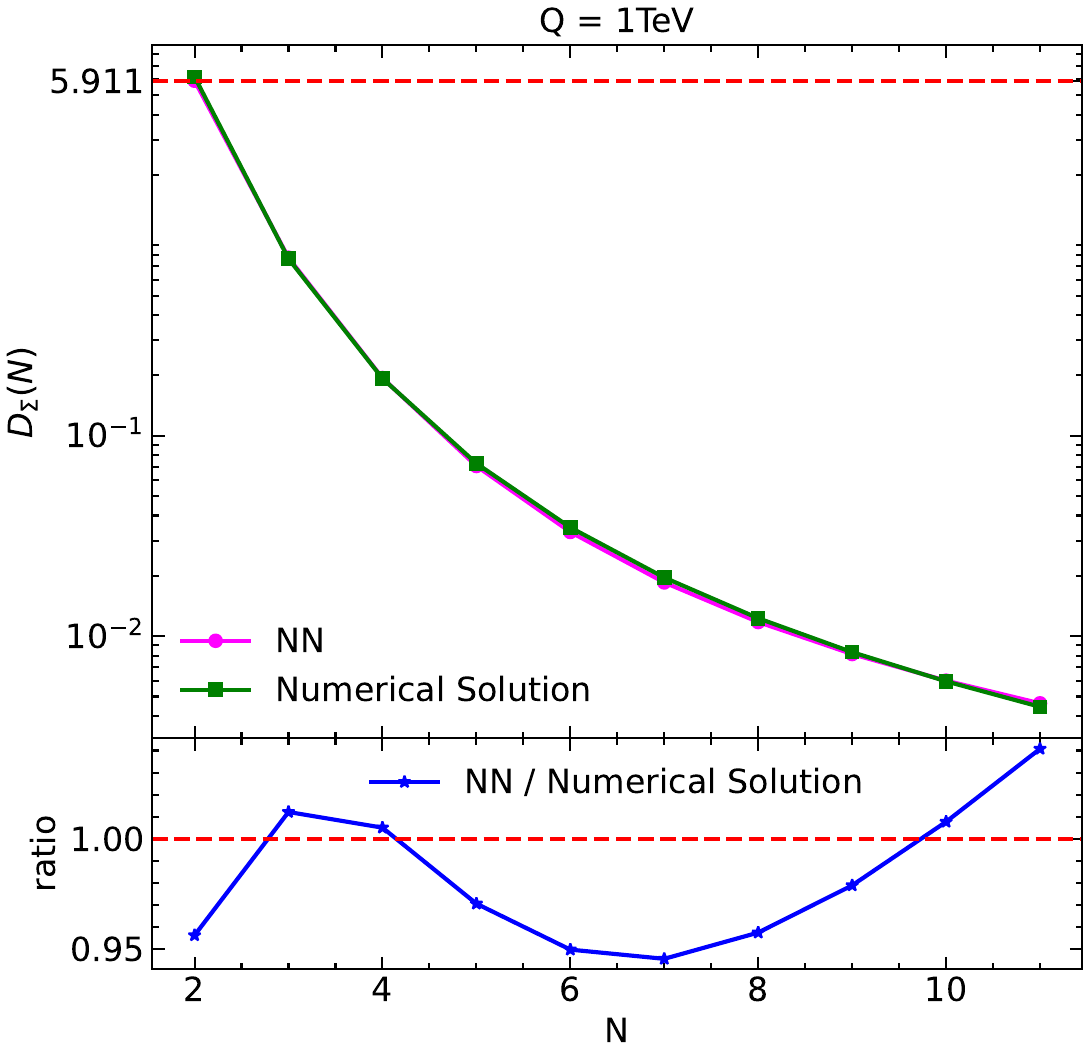}
	\end{minipage}
	\begin{minipage}{0.49\linewidth}
		\centering
		\includegraphics[width=1\linewidth]{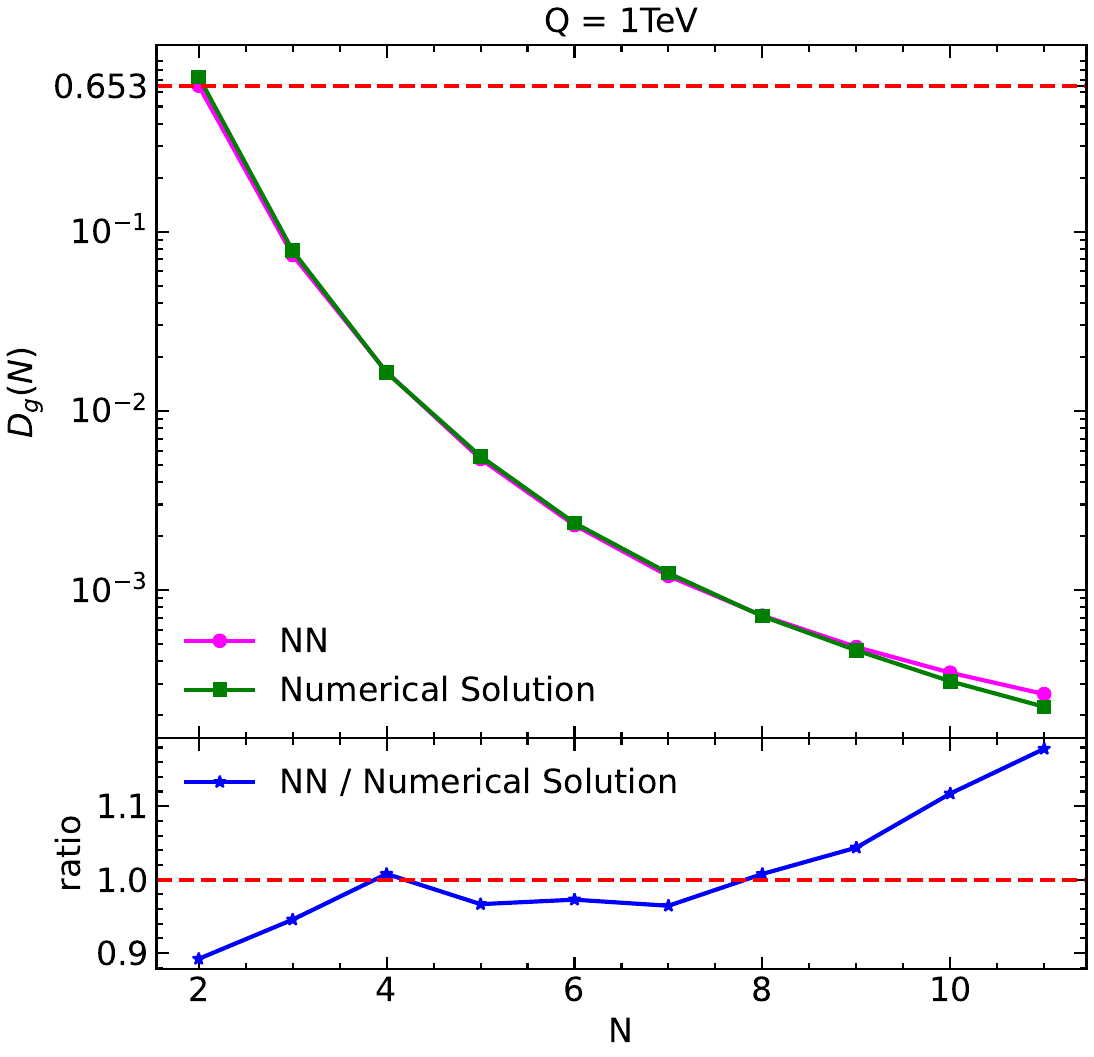}
	\end{minipage}
 \caption{Similar to Fig. \ref{evolution_14} evaluated at $Q=1$ TeV.}
 \label{evolution_1000}
\end{figure*}

Similar to Fig. \ref{evolution_14}, Fig. \ref{evolution_1000} shows FFs evaluated at 1 TeV. It is worth noting that 1 TeV $\notin$ [14 GeV,206 Gev] is outside the energy range of the dataset we used.  The results given by the neural network are still in good agreement with those given by direct evolution, indicating that our constructed neural network satisfies the DGLAP evolution equation. Likewise, the FFs in Fig. \ref{evolution_1000} also satisfy the momentum sum rule, i.e., $D_{\Sigma}^{h^{\pm}}(N = 2,\ Q = 1\ \text{TeV}) = 5.911 < 2n_f$ (five flavors $n_f=5$), $D_g^{h^{\pm}}(2,\ Q = 1\ \text{TeV}) = 0.653 < 1$. 

\subsection{Comparison with AKK08/KRE parameterizations}

\begin{figure*}[htbp]
    \centering
    \includegraphics[width=0.9\linewidth]{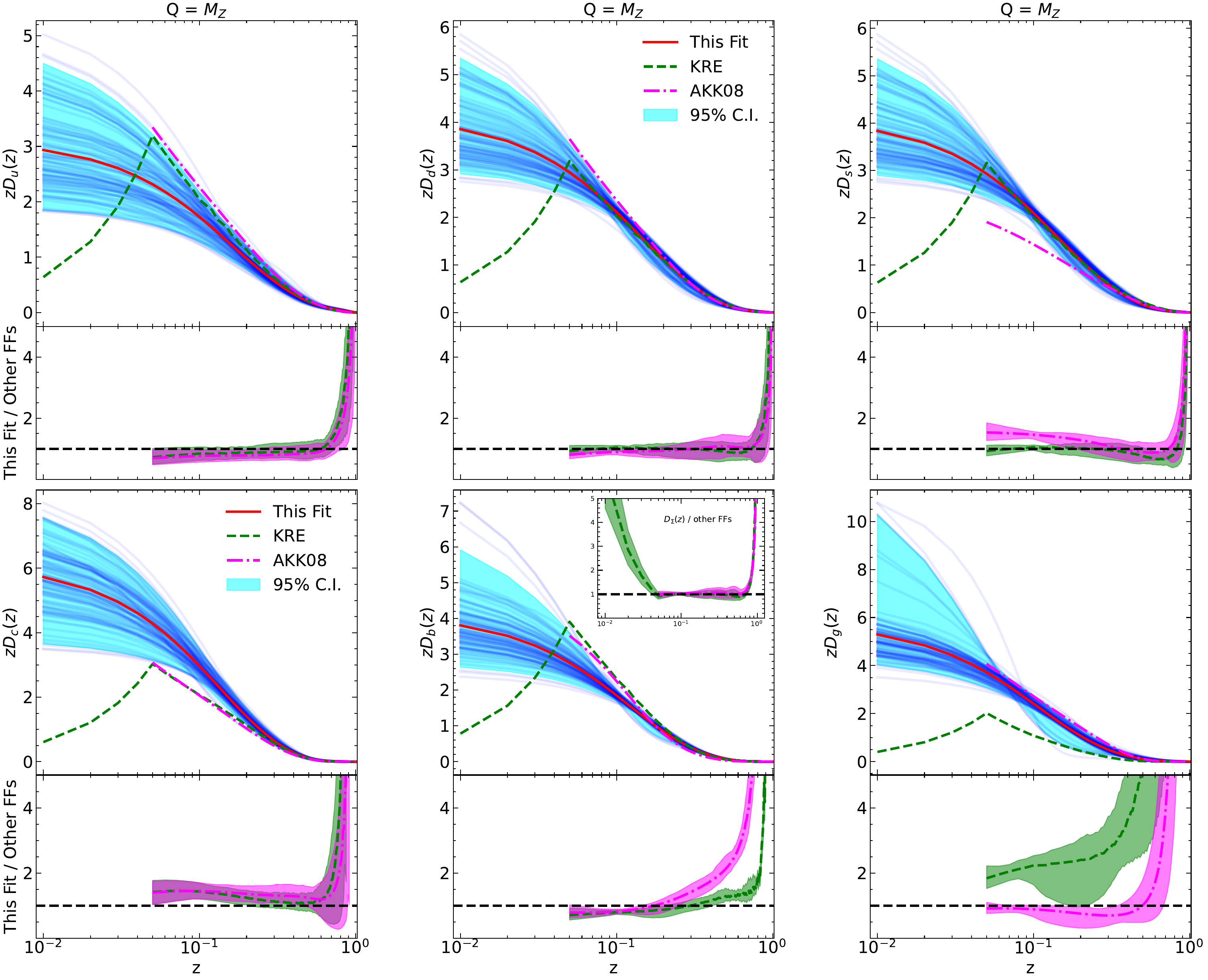}
    \caption{Upper panels: Individual FFs for unidentified charged hadrons $zD_i^{h^{\pm}}(z)$ at $Q$ = $M_Z$. The results of the FFs extracted from 100 random samples (blue curves), together with the mean values and confidence intervals (red curve and cyan bands). The green dashed curve and the magenta dash-dotted curve represent the FFs extracted by KRE \cite{Kretzer:2000yf} and AKK08 \cite{Albino:2008fy}, respectively. Lower panels: ratios of the PINN extracted FFs to the KRE and AKK08 parameterization. The band denotes the confidence interval of 2$\sigma$. We have also plotted the ratio of the singlet quark FFs to the KRE and AKK08 parameterizations in the fifth subplot.}
    \label{FFs_err_anal}
\end{figure*}

Using the Mellin moments from the PINN, we can obtain the FFs as functions of the fractional momentum $z$.
Shown in Fig.~\ref{FFs_err_anal} are FFs of parton $i$ into unidentified charged hadrons at Q = $M_Z$. Blue curves represent the results of the FFs extracted from 100 random samples, the red curve is the mean and the cyan band represents a confidence interval of 2-$\sigma$. For comparison, we also show the KRE (green dashed curve) \cite{Kretzer:2000yf} and AKK08 (magenta dash-dotted curve) \cite{Albino:2008fy} parameterizations. The lower panels display the ratios of our extracted FFs to those of AKK08 and KRE. The KRE imposes a kinematic cut at $z < 0.05$ and freeze the values of FFs at small $z$. AKK08 also applied a kinematic cut at $z=0.05$. In general, the discrepancy among various FFs becomes larger as $z$ approaches 0.01 or 1. The discrepancy between our PINN extracted FFs to those of AKK08 and KRE rises sharply as $z$ nears 1, indicating a slower fall-off of our results in this region. Mathematically, the functional form of our extracted results exhibits a smoother behavior around $z \to 1$.

 Our fit demonstrates agreement with AKK08 and KRE in the singlet combination $D_{\Sigma}^{h^{\pm}}(z,Q)$  in the range $z \in [0.05,0.8]$. However, significant discrepancies are observed in the FFs for $s$, $c$ quarks and gluon FFs, the latter attributed to the data sources included in our analysis. As noted in Refs. \cite{Bertone:2018ecm,DENTERRIA2014615}, the hadron spectrum from RHIC and LHC experiments imposes stronger constraints on gluon FFs. Our analysis exclusively incorporates single-inclusive annihilation (SIA) data, whereas AKK08's analysis included both SIA and proton-proton collision data. Given that collider data directly influences the determination of gluon FF, the observed discrepancy between our gluon FF and those of AKK08/KRE is expected. Interestingly, our extracted gluon FF shows closer agreement with AKK08 compared to KRE. This suggests that our method of extracting parton FFs using PINN may be more accurate than traditional methods that do not incorporate proton-proton collision data. Therefore, including RHIC and LHC experimental data in our analysis could further increase the accuracy of the gluon FF.


\subsection{Closure test of PINN}
One of the most common concerns when employing neural networks (NN) to solve inverse problems is the reliability of the results. Therefore, conducting a closure test for the NN is essential. To this end, we first incorporate the extracted FF into Eq.~(\ref{dx cs}) to calculate the cross sections for single-inclusive (SIA) processes at various center-of-mass energies. Subsequently, the experimental data on the cross-section used in the training of NN is substituted with the calculated values and train the NN again. If the results with such trained NN are approximately consistent with those depicted in Fig. \ref{FFs_err_anal}, this indicates that the NN we have constructed is both reliable and reproducible.

\begin{figure*}[htbp]
    \centering
    \includegraphics[width=0.9\textwidth]{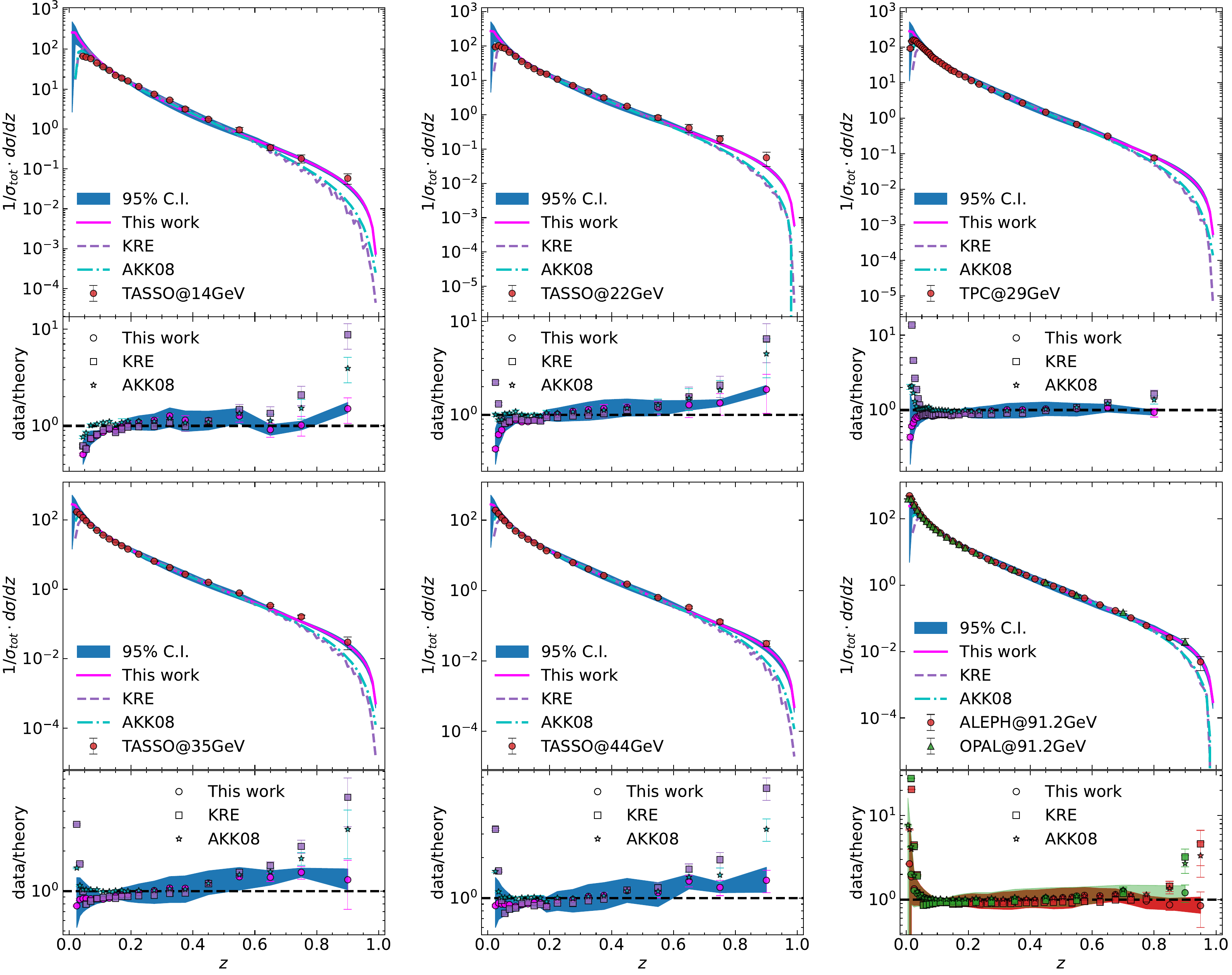}
    \caption{Upper panels: NLO results for single-inclusive charged hadron spectra $e^+e^- \to h^\pm + X$, compared to data from TASSO Collaboration \cite{TASSO:1990cdg}, TPC Collaboration \cite{TPCTwoGamma:1988yjh}, and OPAL Collaboration \cite{OPAL:1998arz} for various c.m.s energies $\sqrt{s}$. Also shown are the results obtained with the KRE \cite{Kretzer:2000yf} and AKK08 \cite{Albino:2008fy} parameterizations. Lower panels: data/theory ratio between our results and the KRE and AKK08 parameterizations. The 95\% confidence intervals in all plots are from our FFs.}
    \label{cross section_1}
\end{figure*}

\begin{figure*}[htbp]
    \centering
    \includegraphics[width=0.9\textwidth]{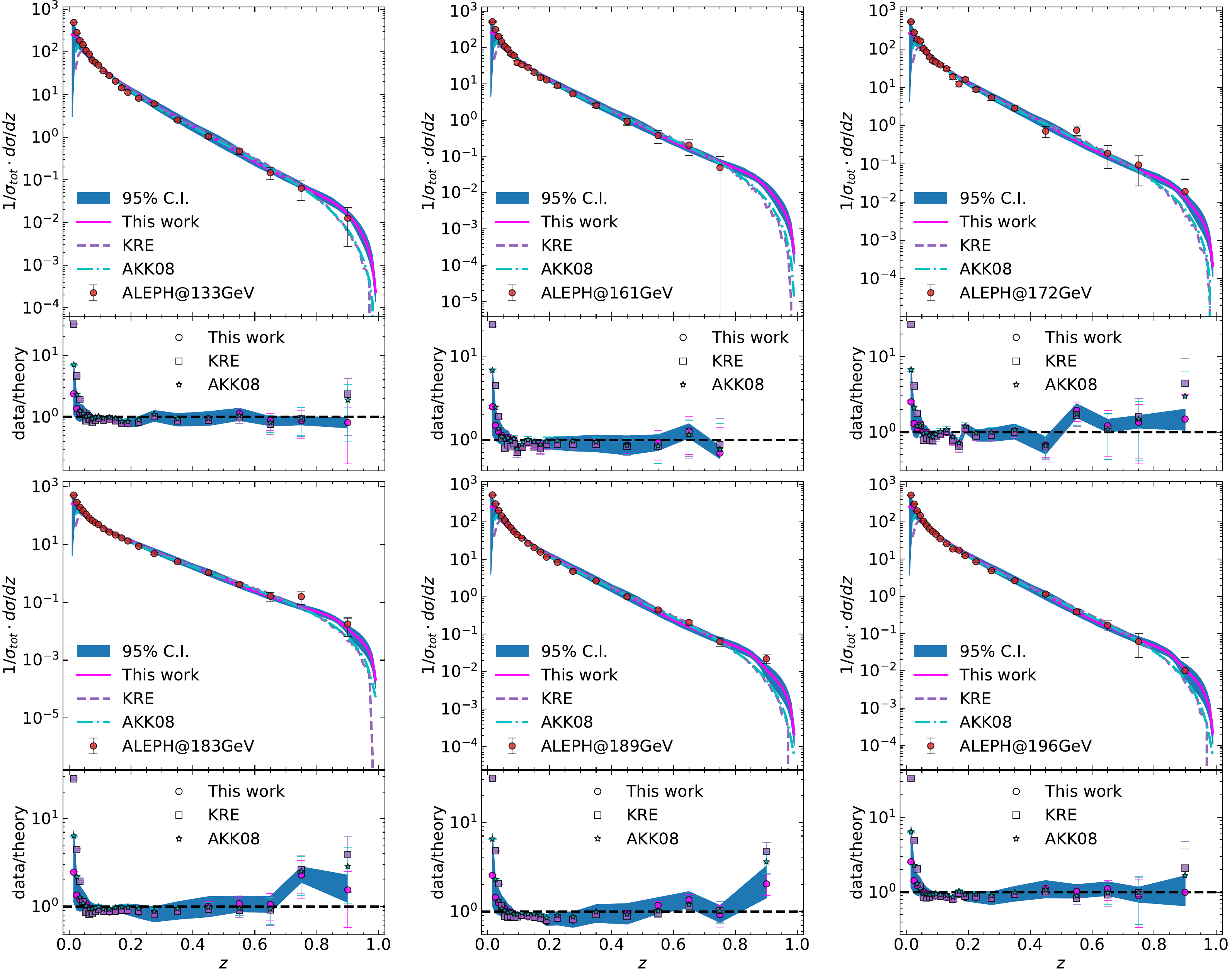}
    \caption{Similar to Fig. \ref{cross section_1} but now comparing to the ALEPH Collaboration \cite{ALEPH:1995njx,ALEPH:2003obs} data at different c.m.s energies $\sqrt{s}$.}
    \label{cross section_2}
\end{figure*}

\begin{figure*}[htbp]
    \centering
    \includegraphics[width=0.9\textwidth]{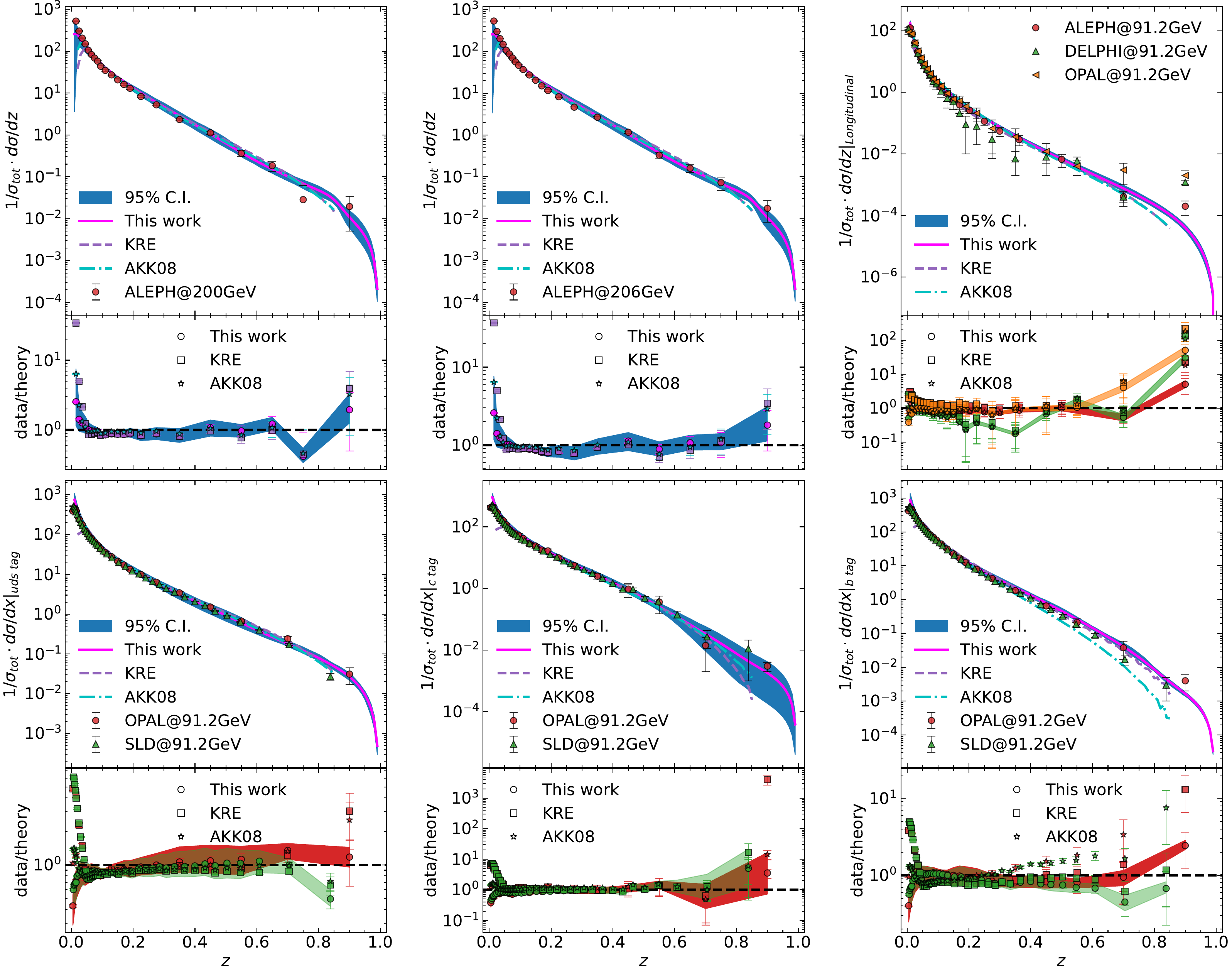}
    \caption{Similar to Fig.~\ref{cross section_1} but compared to the data from ALEPH Collaboration \cite{ALEPH:1995njx,ALEPH:2003obs}, OPAL Collaboration \cite{OPAL:1998arz}, DELPHI Collaboration\cite{DELPHI:1998cgx}, and SLD Collaboration \cite{SLD:2003ogn} at different c.m.s energies $\sqrt{s}$.}
    \label{cross section_3}
\end{figure*}

Figs. \ref{cross section_1} - \ref{cross section_3} illustrate our theoretical results obtained by employing the FFs extracted in this work to calculate the cross section for the SIA process. These figures provide a "data/theory" comparison between the theoretical NLO results and the $e^+e^-$ data set. For comparison, we have also calculated the NLO results using the AKK08 and KRE FFs. The theoretical results computed with our NN, AKK08, and KRE FFs generally agree with the experimental data. However, the theoretical results for AKK08 and KRE are lower than the experimental data in the large $z$ region, while our results agree with the experimental data in the large $z$ region. The primary reason for this discrepancy is that, as previously discussed, our extracted FFs exhibit a gradual decline in the large $z$ region (with $D(z)=0$ strictly at $z=1$), whereas the AKK08 and KRE FFs approach zero before $z \to 1$.

\begin{figure*}[htbp]
   \centering
    \includegraphics[width=0.8\linewidth]{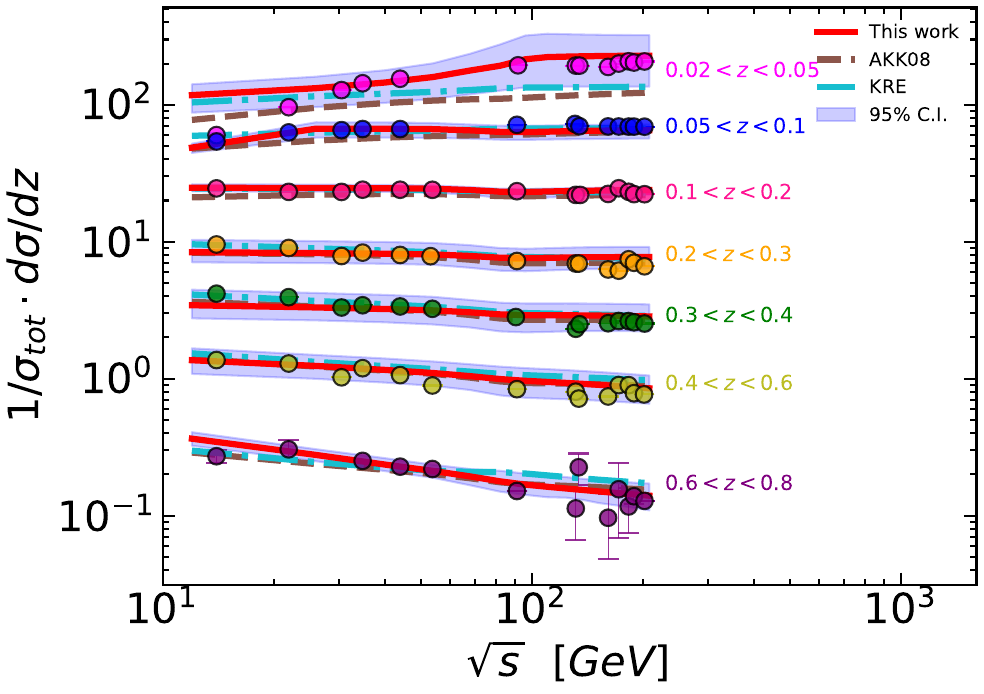}
    \caption{Cross section for $e^+e^- \to h + X$ for charged hadrons for various ranges of $z$ versus $\sqrt{s}$, with experimental data from Particle Data Group \cite{ParticleDataGroup:2018ovx}.}
    \label{cross_section_with_Q}
\end{figure*}

It should be emphasized that Fig.~\ref{Fx and FN} represents the outcome of the NN fitting process, whereas Figs. \ref{cross section_1} - \ref{cross section_3} depict the theoretical results calculated using the extracted FFs. In Fig. \ref{cross_section_with_Q}, we have further shown the variation of the cross section with $Q$ at a fixed $x$ value. The theoretical predictions are found to be consistent with the experimental data. Moreover, as shown in Fig. \ref{cross_section_with_Q}, an increase in the scale induces a scaling violation, characterized by a shift of the $z$-distribution towards lower values, which is consistent with the anticipated scaling violation due to DGLAP evolution.

\begin{figure*}[htbp]
    \centering
    \includegraphics[width=0.9\textwidth]{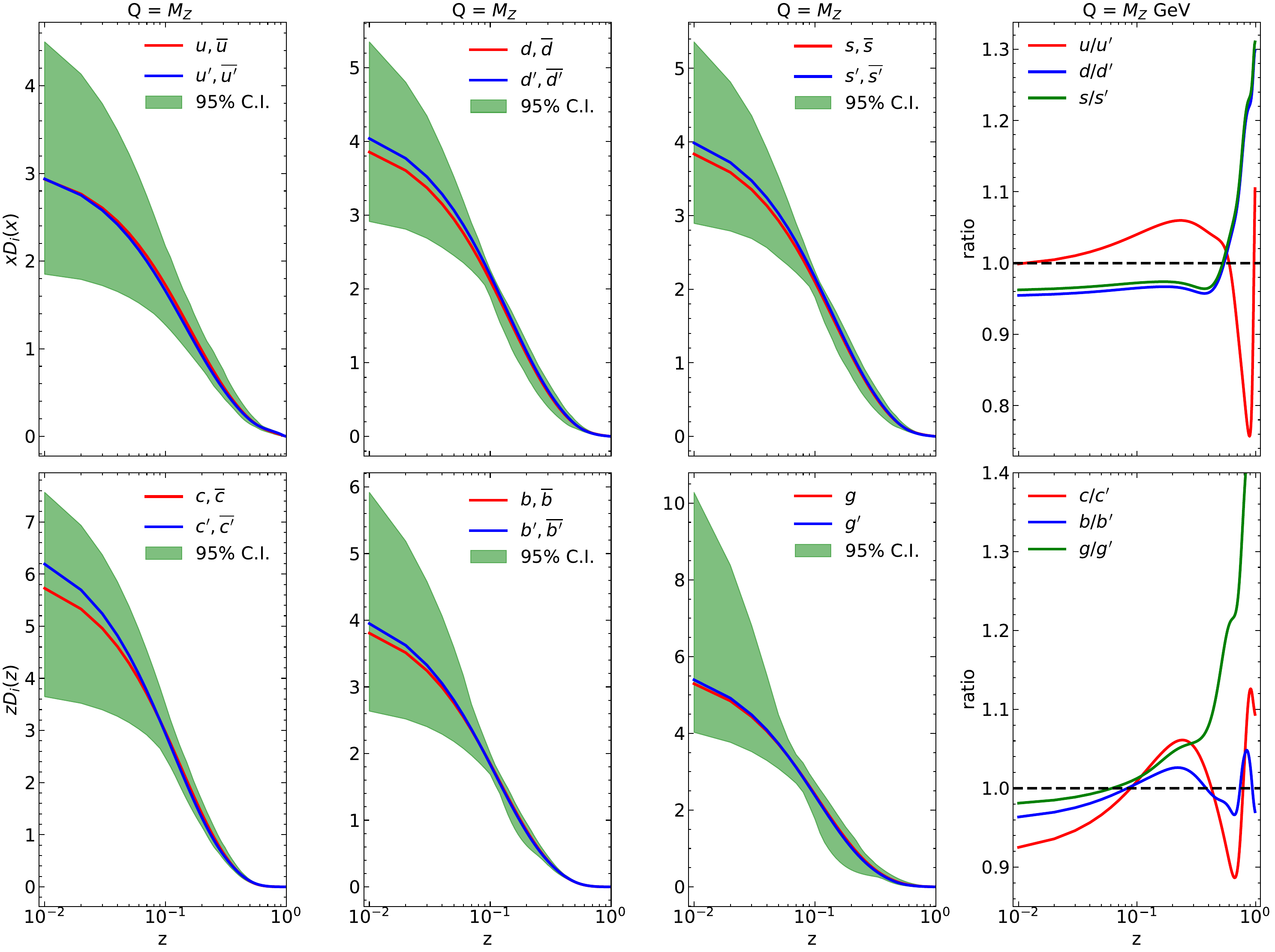}
    \caption{Comparison of the results of the closure test (blue lines) with the original PINN results (red lines) and the ratios of the two (right panels.}
    \label{FFs_test}
\end{figure*}

For the final step of the closure test, theoretical results from Figs. \ref{cross section_1} - \ref{cross section_3} were employed to train the NN. The result is presented in Fig. \ref{FFs_test} as blue lines and compared to the original PINN results (red lines). As one can see the results from the closure test are generally consistent with the original PINN extracted FFs, with the exception in the region near $z \to 1$. The maximum deviation between the two sets of results is 20$\%$ within the interval $z \in [0.01,0.8]$, and are within the confidence intervals depicted in Fig. \ref{FFs_test}. This demonstrates that the NN we have constructed is stable and reproducible.

\subsection{Hadron spectra in $pp(\bar{p})$ collisions}

To further verify the PINN extracted FFs in our analysis, we use them to compute the single inclusive hadron spectra and compare to the experimental data in $p+p(\bar{p}) \to h^{\pm}X$ at RHIC and LHC energies. 

\begin{figure*}[htbp]
    \centering
    \includegraphics[width=0.9\linewidth]{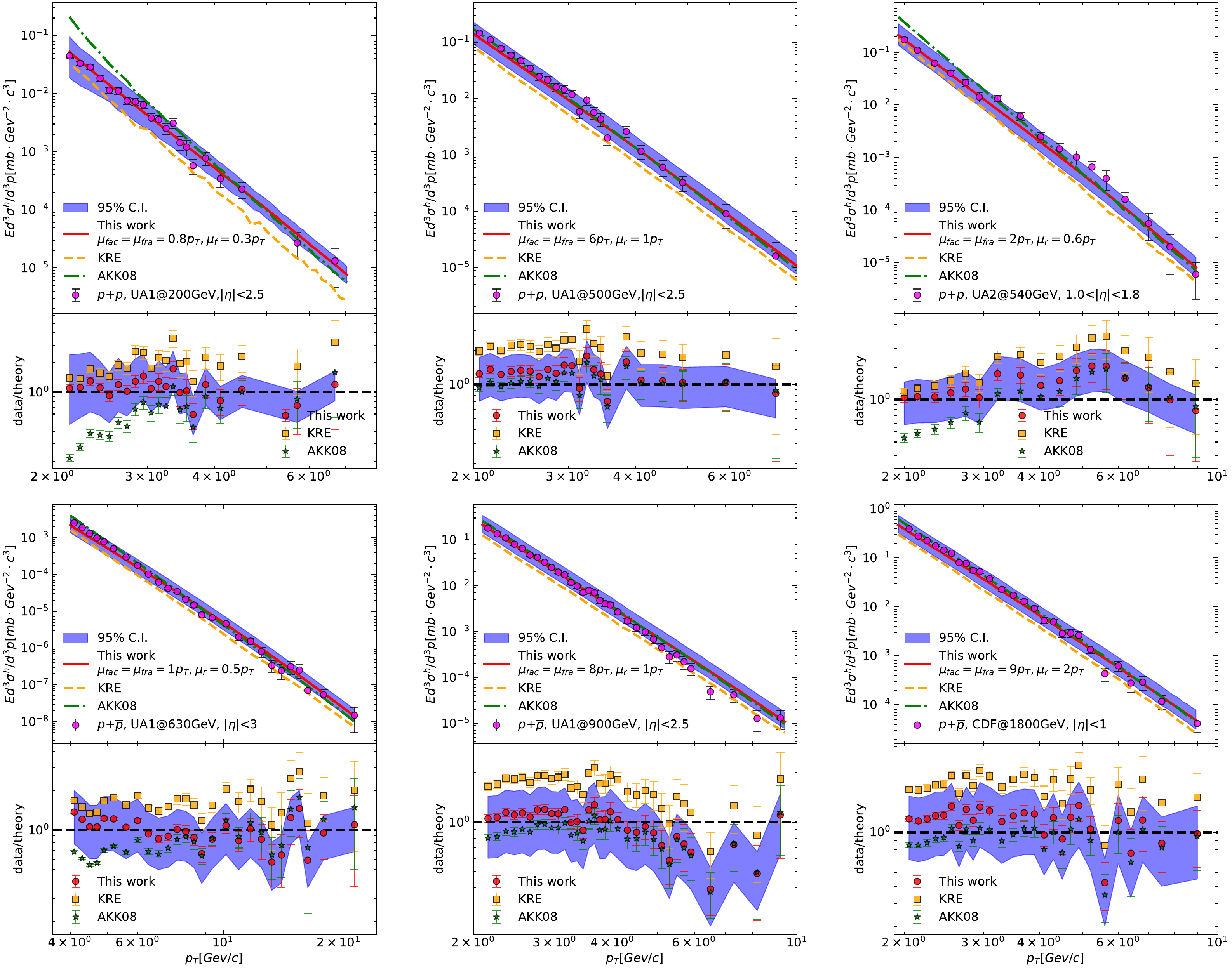}
    \caption{Upper panels: NLO results for single-inclusive charged hadron production $p\bar{p} \to h^\pm + X$ with PINN extracted FFS as compared to data from UA1 \cite{UA1:1989bou,Bocquet:1995jr}, UA2 \cite{UA2:1984ida} and CDF \cite{CDF:1988evs} as well as NLO results with the KRE \cite{Kretzer:2000yf} and AKK08 \cite{Albino:2008fy} parametrization of FFs. Lower panels: data/theory ratio for our results and the KRE and AKK08 parameterizations. The 95\% confidence intervals in all plots are from our FFs.}
    \label{UA1}
\end{figure*}

\begin{figure*}[htbp]
    \centering
    \includegraphics[width=0.9\linewidth]{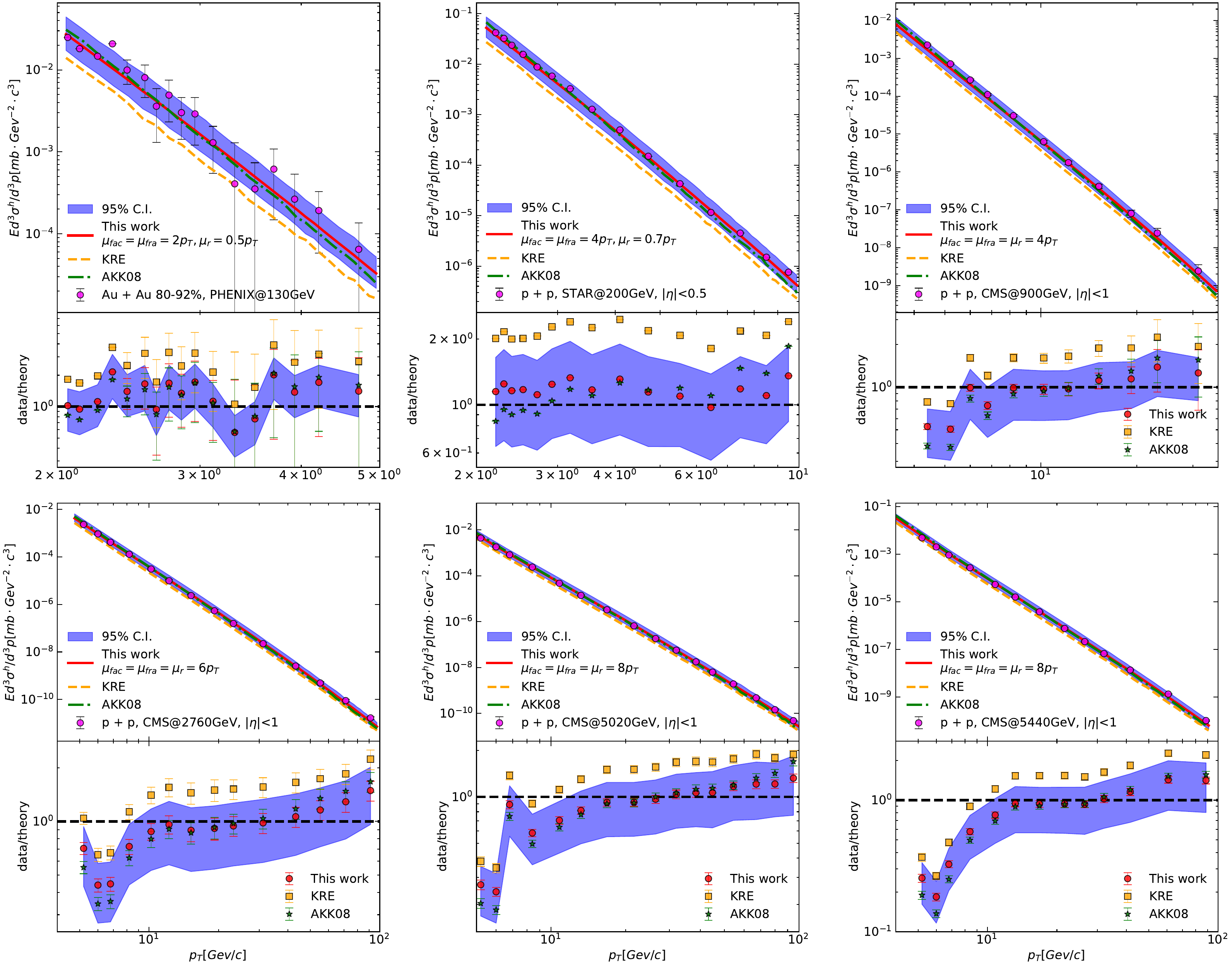}
    \caption{Upper panels: NLO results with PINN extracted FFs for single-inclusive charged hadron production $p+p \to h^\pm + X$ as compared to data from the STAR \cite{STAR:2003fka} PHENIX \cite{PHENIX:2002diz} and CMS \cite{CMS:2011mry,CMS:2012aa,CMS:2016xef,CMS:2018yyx} as well as results with the KRE \cite{Kretzer:2000yf} and AKK08 \cite{Albino:2008fy} parameterizations of FFS. Lower panels: data/theory ratio between our results and the KRE and AKK08 parameterizations. The 95\% confidence intervals in all plots are from our FFs.}
    \label{CMS}
\end{figure*}

\begin{figure*}[htbp]
    \centering
    \includegraphics[width=0.9\linewidth]{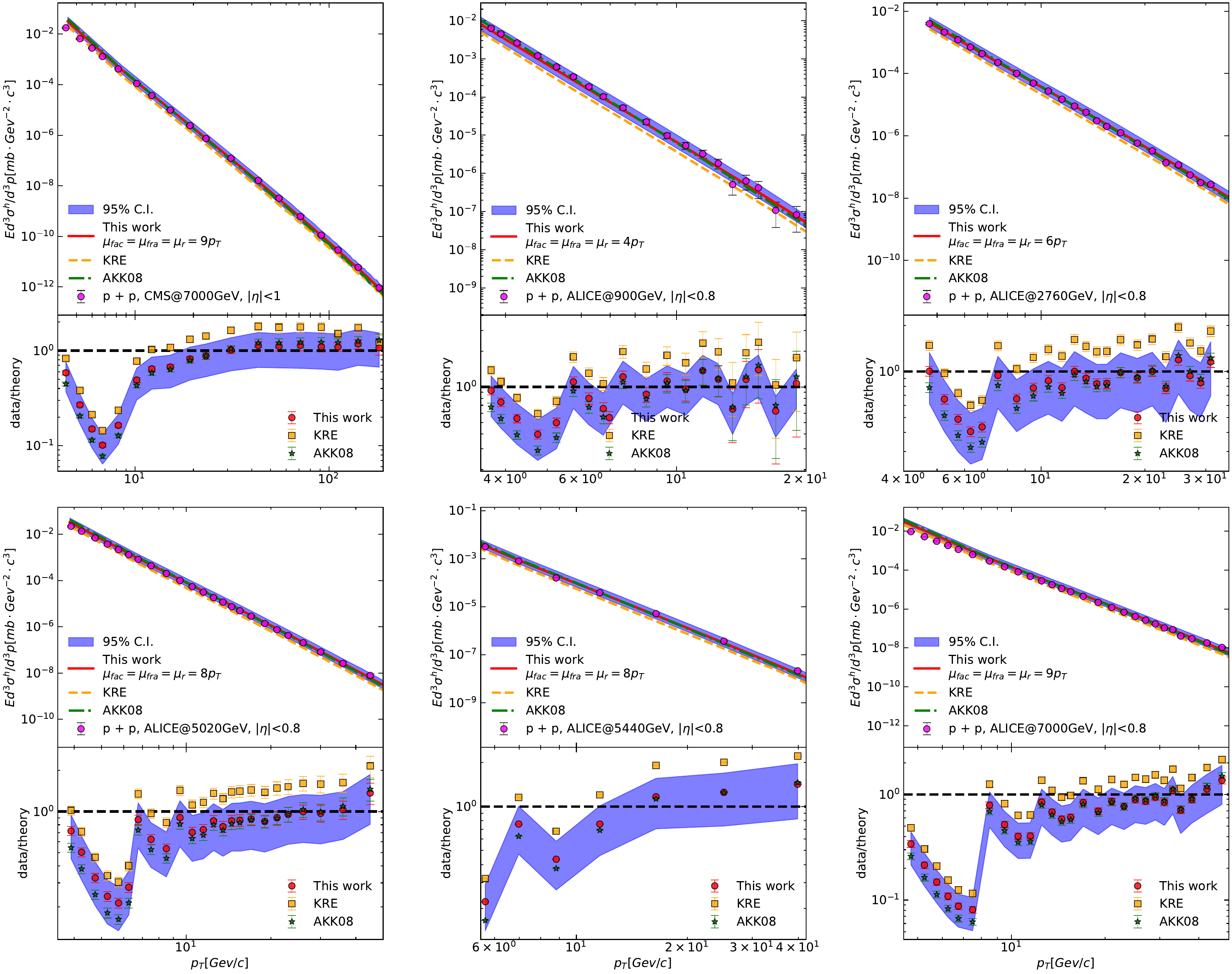}
    \caption{Similar to Fig. \ref{CMS} except comparing to the ALICE \cite{ALICE:2013txf,ALICE:2014nqx,ALICE:2018hza} data LHC.}
    \label{ALICE}
\end{figure*}

Figs.~\ref{UA1}, \ref{CMS} and \ref{ALICE} show the comparisons between our NLO calculations of the single-inclusive cross sections($p + p(\bar{p}) \to h^{\pm} + X$) and the experimental data from UA1 \cite{UA1:1989bou,Bocquet:1995jr}, UA2 \cite{UA2:1984ida} and CDF \cite{CDF:1988evs},  ALICE \cite{ALICE:2013txf,ALICE:2014nqx,ALICE:2018hza}, STAR \cite{STAR:2003fka}, PHENIX \cite{PHENIX:2002diz}, CMS \cite{CMS:2011mry,CMS:2012aa,CMS:2016xef,CMS:2018yyx}, respectively at RHIC and LHC collision energies. The data-to theory ratios are plotted in the lower panels. 
Note that in Fig. \ref{CMS}, where experimental data for $pp$ collisions at $\sqrt{s}$ = 130 GeV are unavailable, we employ peripheral Au+Au collision data (80-92\% centrality) and the Glauber model of multiple scattering ${d\sigma^{AA} \over dp} = \langle N_{binary}\rangle{d\sigma^{pp} \over {dp}}$ to approximate the cross section in $pp$ collisions. This approximation relies on the assumption that final-state parton energy loss is negligible in these peripheral collisions. In addition, we acknowledge that initial-state Cold Nuclear Matter (CNM) effects, such as the nuclear modification of PDFs (nPDFs), are treated as minimal, though they theoretically introduce additional systematic uncertainty.
For comparison, we also provide theoretical predictions using the AKK08 and KRE parameterizations of FFs, all evaluated at the same factorization scale.


It is evident that NLO pQCD results with our PINN extracted FFs generally can describe well the experimental data at large $p_T>10$ GeV from RHIC to LHC energies. They also agree with those of AKK08 FFs, while those of KRE lie on the boundary of the confidence interval of our results. It is important to note that there are better factorization scales than our chosen scale for AKK08/KRE, which allows for adjustments to better fit the experimental data using KRE FFs. However, our primary focus here is not on whether the results computed with the FFs match the experimental data but rather on the differences between the different FFs with the same scale. As shown in Figs.~\ref{UA1} - \ref{ALICE}, the shapes calculated by these three sets of different FFs are consistent with each other. This indicates that our neural network (NN) FFs conform to our expectations.

\section{SUMMARY}
\label{FIVE}
In this study, we present a novel technique to determine the unidentified charged hadron fragmentation functions (FFs) at next-to-leading order (NLO) in perturbative QCD. Our analysis incorporates flavor-untagged and tagged single-inclusive annihilation (SIA) data in $e^+e^-$ annihilation from various experiments, including ALEPH \cite{ALEPH:1995njx,ALEPH:2003obs}, TASSO \cite{TASSO:1990cdg}, TPC \cite{TPCTwoGamma:1988yjh}, OPAL \cite{OPAL:1998arz}, DELPHI \cite{DELPHI:1998cgx}, and SLD \cite{SLD:2003ogn}. The extraction of FFs is achieved using Physics Informed Neural Networks (PINN), a deep learning approach that integrates both the distributional properties of training data like traditional neural networks and the physical laws such as the DGLAP evolution equations for FFs. We list our key findings, limitation and future work as follows:

\textbf{Non-Parametric Form of FFs}: Unlike traditional methods that assume a specific parametric form for FFs, our approach does not impose any such assumptions. This non-parametric approach provides a more robust and unbiased estimation of FFs, potentially uncovering new features that parametric methods might overlook. 
Our method allows the extracted non-parametric FFs to naturally satisfy the DGLAP evolution equations, as the PINN technique embeds these equations directly into the neural network. 

\textbf{Inverse Mellin Transform and DGLAP Evolution}: The neural network constructed in this work is capable of solving the inverse Mellin transform and the DGLAP evolution equations. Compared to traditional methods, the neural network demonstrates superior performance in solving the inverse Mellin transform and the DGLAP evolution equations, as PINN can handle complex boundary conditions and can be computed without the need of a mesh.

\textbf{Hadron Spectrum and FF Universality}: NLO pQCD calculations using the FFs extracted in this work can describe well the hadron spectrum in $p+ p(\bar{p}) \to h^{\pm} + X$ at RHIC and LHC energies.
This not only validates the numerical accuracy of our FFs but also underscores their universality across different energy regimes and collision systems.

\textbf{Constraints on Gluon and Heavy Quark FFs}: The absence of experimental data from electron-proton ($ep$) and proton-proton ($pp$) collisions in our analysis results in weaker constraints on gluon and heavy quark FFs. Consequently, our extracted gluon and heavy quark FFs differ from those in the AKK08/KRE parameterization. However, the singlet quark FFs remain consistent with AKK08/KRE. 
Compared to the SIDIS and $p+p$ experimental data, the $e^+e^-$ annihilation exhibits weaker constraints on the gluon FF at the NLO and NNLO order, such that the gluon FF is not significantly constrained using only SIA data. However, SIA data provides a strong constraint on $D_q^{h^{\pm}} + D_{\bar{q}}^{h^{\pm}}$,  thereby effectively constraining the singlet quark FFs.

\textbf{Future work}: In the future we will incorporate data from semi-inclusive deep inelastic scattering (SIDIS) and hadron collider experiments, including $ep$ and $pp$ data. As discussed in this analysis, including these datasets will strengthen the constraints on FFs, particularly for gluon and heavy quark FFs, leading to more precise and comprehensive fragmentation functions. This extension will further validate the universality and robustness of the FFs extracted using the PINN approach.

Our study demonstrates the potential of PINN in accurately determining FFs without the need for parametric assumptions, while naturally satisfying physical constraints such as the DGLAP evolution equations and works well across different colliding energies  and collision systems. The inclusion of additional experimental data in future analyses will enhance the precision and applicability of these FFs, and also allow for the consideration of a modified DGLAP evolution equations \cite{Guo:2000nz,Wang:2001ifa,Zhang:2004qm} due to medium-induced effects. Such consideration could enable the study of jet quenching effects \cite{Xie:2022ght,Xie:2022fak,Xie:2024xbn} in high-energy heavy-ion collisions.



\noindent
{\bf{Acknowledgements}}
This work has been supported by by the NSFC under grant Nos.\ 11935007, 12075098, 12535010 and No.\ 12435009, and 
by Guangdong Major Project of Basic and Applied Basic Research under Grant No. 2020B030103008.

\noindent
{\bf{Data Availability Statement}} This manuscript has no associated data or the data will not be deposited. [Authors’ comment: Data points used to draw the plots and predictions for the future measurements can be obtained through email upon request.]

\clearpage

\appendix
\onecolumngrid  

\makeatletter
\@addtoreset{equation}{section}
\makeatother
\renewcommand{\theequation}{\Alph{section}\arabic{equation}}
\setcounter{equation}{0}

\section{\texorpdfstring{$\MSbar$}{MSbar} Coefficient Functions}
\label{app:coeff_functions}

In this appendix we give the coefficient functions in $z$-space and Mellin moment space at NLO,
respectively~\cite{Nason:1993xx,Altarelli:1979kv,Furmanski:1981cw}.

\subsection*{$z$-space}

\begin{subequations}\label{eq:coeff-z}
\begin{align}
C_T^q(z)
&= \sigma_0^q(s)\Bigg\{\delta(1-z)
+\frac{2\alpha_s}{3\pi}\Bigg[
(1+z^2)\left[\frac{\ln(1-z)}{1-z}\right]_+
-\frac{3}{2}\left[\frac{1}{1-z}\right]_+ \notag\\
&\qquad
+2\frac{1+z^2}{1-z}\ln z
+\frac{3}{2}(1-z)
+\left(\frac{2}{3}\pi^2-\frac{9}{2}\right)\delta(1-z)
+\Lmu\left[\frac{1+z^2}{1-z}\right]_+
\Bigg]\Bigg\}\sigma_0^q(s),
\\
C_T^g(z)
&= \frac{4\as}{3\pi}\Bigg[
\frac{1+(1-z)^2}{z}\Big(\ln(1-z)+2\ln z + \Lmu\Big)
-2\frac{1-z}{z}
\Bigg]\sum_q \sigma_0^q(s),
\\
C_L^q(z)
&= \frac{2\as}{3\pi}\sigma_0^q(s),
\\
C_L^g(z)
&= \frac{8\as}{3\pi}\frac{1-z}{z}\sum_q \sigma_0^q(s).
\end{align}
\end{subequations}

\subsection*{Mellin moment space}

\begin{subequations}\label{eq:coeff-N}
\begin{align}
C_T^q(N)
&= \sigma_{0}^q(s)\Bigg\{1+\frac{2\as}{3\pi}\Bigg[
5S_2(N)+S_1^2(N)-\frac{2}{N^2}-\frac{9}{2}
+S_1(N)\frac{3N^2+3N-2}{2N(N+1)}
+\frac{9}{2(N+1)^3}
\Bigg]\Bigg\},
\\
C_T^g(N)
&= \frac{4\as}{3\pi}\Bigg[
-S_1(N)\frac{N^2+N+2}{(N-1)N(N+1)}
-\frac{4}{(N-1)^2}
+\frac{4}{N^2}
-\frac{3}{(N+1)^2}
\Bigg]\sum_q \sigma_0^q(s),
\\
C_L^q(N)
&= \frac{2\as}{3\pi}\frac{1}{N}\sigma_0^q(s),
\\
C_L^g(N)
&= \frac{2\as}{3\pi}\frac{4}{N(N-1)}\sum_q \sigma_0^q(s).
\end{align}
\end{subequations}

\noindent where
\begin{align}
S_1(N) &= \gamma_E + \psi(N+1),\\
S_2(N) &= \frac{\pi^2}{6} - \psi'(N+1),
\end{align}
and
\begin{equation}
\psi^{(m)}(N)=\frac{d^{m+1}}{dN^{m+1}}\ln\Gamma(N).
\end{equation}

\twocolumngrid
\clearpage
\normalem
\bibliography{ref.bib}

\end{document}